\documentclass[letterpaper,12pt]{JHEP3}
\pdfoutput=1
\usepackage{amsmath}
\usepackage{amssymb}
\usepackage{bbm}
\usepackage{subfigure} 
\usepackage{epsfig}
\usepackage{yfonts}
\usepackage{graphicx}

\newcommand{\labell}[1]{\label{#1}}

\newcommand{\be}{\begin{equation}}
\newcommand{\ee}{\end{equation}}
\newcommand{\bea}{\begin{eqnarray}}
\newcommand{\eea}{\end{eqnarray}}
\newcommand{\ba}{\begin{eqnarray}}
\newcommand{\ea}{\end{eqnarray}}

\newcommand{\beq}{\begin{equation}}
\newcommand{\eeq}{\end{equation}}
\newcommand{\beqa}{\begin{eqnarray}}
\newcommand{\eeqa}{\end{eqnarray}}
\newcommand{\beqar}{\begin{eqnarray*}}
\newcommand{\eeqar}{\end{eqnarray*}}

\newcommand{\reef}[1]{(\ref{#1})}
\newcommand{\ssc}{\scriptscriptstyle}
\newcommand{\eg}{{\it e.g.,}\ }
\newcommand{\ie}{{\it i.e.,}\ }

\newcommand{\mt}[1]{\textrm{\tiny #1}}

\newcommand{\veps}{\varepsilon}

\newcommand{\R}{\mathcal{R}}

\newcommand{\D}{\mathcal{D}}
\newcommand{\E}{\mathcal{E}}

\newcommand{\cO}{\mathcal{O}}
\newcommand{\cM}{\mathcal{M}}
\newcommand{\hH}{\mathcal{H}} 

\newcommand{\tr}{{\tilde \rho}}
\newcommand{\hr}{{\hat r}}
\newcommand{\vrho}{\varrho}
\newcommand{\trho}{{\tilde\varrho}}
\newcommand{\ttau}{{\tilde \tau}}

\newcommand{\ads}{a_d^*}

\newcommand{\too}[1]{\mathrel{\mathop \to_{\scriptscriptstyle{#1}}}{}\!\!}

\preprint{arXiv:1102.0440 [hep-th]}
\title{Towards a derivation of holographic entanglement entropy}
\author{Horacio Casini,$^{1}$ Marina Huerta$^{1}$ and Robert C. Myers$^{2}$\\
\it $^1$Centro At\'omico Bariloche and Instituto Balseiro\\
\ 8400-S.C. de Bariloche, R\'\i o Negro, Argentina\\
$^2$Perimeter Institute for Theoretical Physics\\
\ Waterloo, Ontario N2L 2Y5, Canada}

\abstract{We provide a derivation of holographic entanglement entropy
for spherical entangling surfaces. Our construction relies on conformally
mapping the boundary CFT to a hyperbolic geometry and observing that the
vacuum state is mapped to a thermal state in the latter geometry. Hence the
conformal transformation maps the entanglement entropy to the thermodynamic
entropy of this thermal state. The AdS/CFT dictionary allows us to calculate
this thermodynamic entropy as the horizon entropy of a certain topological black
hole. In even dimensions, we also demonstrate that the universal contribution
to the entanglement entropy is given by $A$-type trace anomaly for any CFT,
without reference to holography.}


\begin{document}

\section{Introduction} \label{intro}

Entanglement entropy has become an important quantity for the study of
quantum matter. It allows one to distinguish new topological phases and
characterize critical points, \eg \cite{wenx,cardy0,fradkin}.
Entanglement entropy has also been considered in discussions of
holographic descriptions of quantum gravity, in particular, for the
AdS/CFT correspondence \cite{ryut0,ryut1}. In this context, as well as
characterizing new properties of holographic field theories, \eg
\cite{igor0}, it has been suggested that entanglement entropy may
provide new insights into the quantum structure of spacetime
\cite{mvr}.

The proposal \cite{ryut0,ryut1} of how to calculate holographic
entanglement entropy is both simple and elegant. In the boundary field
theory, one would begin by chosing a particular spatial region $V$. The
entanglement entropy between $V$ and its complement $\bar V$ would then
be the von Neumann entropy of the density matrix which results upon
integrating out the field theory degrees of freedom in $\bar V$. In the
holographic calculation which is proposed to yield the same entropy,
one considers bulk surfaces $m$ which are "homologous"
\cite{head1,fur0} to the region $V$ in the boundary (in particular
$\partial m=\partial V$). Then one extremizes the area\footnote{If the
calculation is done in a Minkowski signature background, the extremal
area is only a saddle point. However, if one first Wick rotates to
Euclidean signature, the extremal surface will yield the minimal area.
In either case, the area must be suitably regulated to produce a finite
answer. Further note that for a $d$-dimensional boundary theory, the
bulk has $d+1$ dimensions while the surface $m$ has $d-1$ dimensions.
We are using `area' to denote the ($d-1$)-dimensional volume of $m$.}
of $m$ to calculate the entanglement entropy:
 \be
S(V) =  \mathrel{\mathop {\rm ext}_{\scriptscriptstyle{m\sim V}}
{}\!\!} \left[\frac{A(m)}{4G_\mt{N}}\right]\,.
 \labell{define}
 \ee
An implicit assumption in eq.~\reef{define} is that the bulk physics is
described by (classical) Einstein gravity. Hence we might note the
similarity between this expression \reef{define} and that for black
hole entropy. While this proposal to calculate the holographic
entanglement entropy passes a variety of consistency tests, \eg see
\cite{head1,ryut2,mishanet}, there is no concrete construction which
allows one to derive this holographic formula \reef{define}.

A standard approach to calculating entanglement entropy in field theory
makes use of the replica trick \cite{cardy0,callan}. This approach
begins by calculating the partition function on an $n$-fold cover of
the background geometry where a cut is introduced throughout the
exterior region $\bar V$. If one were to apply this construction for
the boundary CFT in holographic framework, one would naturally produce
a conical singularity in the bulk geometry with an angular excess of
$2\pi(n-1)$. However, without a full understanding of string theory or
quantum gravity in the AdS bulk, we do not understand how to resolve
the resulting conical curvature singularity and so it is not really
possible work with this bulk geometry in a controlled way, \eg it is
not possible to properly evaluate the saddle-point action in the
gravity theory. This issue was emphasized in \cite{head1} in critiquing
the attempted derivation of \cite{fur0}. Further, it was demonstrated
there that the approach of \cite{fur0} leads to incorrect results in
holographic calculations of Renyi entropies.

Hence the replica trick does not seem a useful starting point in
considering a derivation of holographic entanglement entropy. An
interesting derivation of the holographic entanglement entropy for a
specific geometry was recently presented in \cite{cthem,cthem2}. In
particular, the boundary CFT was placed on an $R\times S^{d-1}$
background and the entanglement entropy was calculated for an
entangling surface that divided the sphere into two halves. It was
argued that the holographic entanglement entropy was given by the
horizon entropy of a certain topological black hole whose horizon
divided the bulk geometry in half. In this paper, we clarify this
construction and in doing so extend the derivation to more general
spherical entangling surfaces.

In section \ref{cft}, we set aside holography and begin with a
discussion of the entanglement entropy for spherical entangling
surfaces for a general conformal field theory (CFT). In particular, we
use conformal transformations to map the causal development of the
interior region $V$ to a new geometry which is the direct product of
time with the hyperbolic plane $H^{d-1}$. Further, we demonstrate that
if the CFT began in the vacuum in the original space, in this new
geometry, we have a thermal bath whose temperature is controlled by the
size of the sphere. Hence the density matrix describing the CFT inside
the sphere becomes a thermal density matrix on the hyperbolic geometry
(up to a unitary transformation which, of course, preserves the
entropy). Hence the entanglement entropy across the sphere becomes the
thermodynamic entropy in the second space. This construction was, in
part, inspired by the appearance of a hyperbolic space in the
calculations of entanglement entropies across spheres for a free scalar
in \cite{casini1} and it generalizes some older observations in
ref.~\cite{candow}, again for free field theories.

While the results outlined above are quite general, in particular
applying for any arbitrary CFT, it may seem this discussion only
replaces one difficult problem, the calculation of the entanglement
entropy, with another equally difficult problem, the calculation of the
thermal entropy in a hyperbolic space. However, we are particularly
interested in applying this result to the AdS/CFT correspondence. In
this framework, the standard holographic dictionary suggests that the
thermal state in the boundary CFT is dual to a black hole in the bulk
gravity theory. Hence in section \ref{ads}, we identify the
corresponding black hole dual to the thermal bath on the hyperbolic
geometry. The latter turns out to be a certain topological black hole
with a hyperbolic horizon, but also one which is simply a hyperbolic
foliation of the empty AdS spacetime. Hence with this bulk
interpretation, we are able to compute the entanglement entropy as the
horizon entropy of the black hole. This construction is not limited to
having Einstein gravity in the bulk and so in fact, a general result is
presented for any gravitational theory.

A generalization of the replica trick \cite{callan} has been applied to
relate entanglement entropy to the trace anomaly for CFT's in an even
number of spacetime dimensions \cite{ryut1,solo1}. To be precise, with
a general entangling surface, the universal coefficient of the
logarithmic contribution to the entanglement entropy is given by some
linear combination of the central charges appearing in the trace
anomaly for any CFT with even $d$. The latter can be written as
\cite{duff}
\be \langle\,T^\mu{}_\mu\,\rangle = \sum B_n\,I_n -2\,(-)^{d/2}A\,
E_d\,. \labell{traza} \ee
where $E_d$ is the Euler density in $d$ dimensions and $I_n$ are the
independent Weyl invariants of weight $-d$.\footnote{Note that we have
discarded a scheme dependent total derivative on the right-hand side of
eq.~\reef{traza}. For more details on our conventions here, see
\cite{cthem2}.} A result of our holographic analysis in section
\ref{ads} is that the linear combination determining the universal
contribution for a spherical entangling surface reduces to be simply
the coefficient of the $A$-type trace anomaly. In section \ref{CFT2},
we obtain the same result for an arbitrary CFT, without reference to
holography. There, our approach again relies on using a conformal
mapping of the geometry in the entanglement entropy calculation.
However, in this case, our conformal mapping takes the CFT to (the
static patch of) de Sitter space, where the state is again thermal, and
the entropy is again interpreted as ordinary thermodynamical entropy.

A connection of the coefficient
of the logarithmic term in the entanglement entropy of a sphere with
the $A$-type anomaly was also found previously for $d=4$ by
\cite{solo1}. Further, this connection was also noted for free scalar
fields in any even dimension in \cite{stud3}. Let us note that there
have been a number of other recent works related to the entanglement
entropy of free fields for a spherical entangling surface
\cite{casini1,dowker1,solo2,dowker2}.

\section{The CFT story} \label{cft}

In this section, we describe the entanglement entropy of any general
conformal field theory for spherical entangling surfaces in certain
background geometries. We emphasize that the discussion here is made
purely in the context of quantum field theory (QFT), without reference
to holography. However, the results found here will set the stage of a
holographic calculation of the same entanglement entropy using the
AdS/CFT correspondence in section \ref{ads}. In particular, we consider
here a $d$-dimensional CFT in Minkowski space $R^{1,d-1}$ and examine
the entanglement entropy across a spherical surface $S^{d-2}$. As
described above, we show that the causal development of the region
inside the $S^{d-2}$, which we denote $\D$, can be mapped to a space
$\hH\equiv R\times H^{d-1}$. Further we show that the vacuum
correlators in $\D$ are conformally mapped to thermal correlators in
the space $\hH$. Hence we are able to show that the density matrix on
$\D$ is given as a Gibbs state of a local operator which is just
constructed by conformally mapping the (curved space) Hamiltonian of
the CFT in $\hH$ back to $\D$. With this construction, the entanglement
entropy across the sphere becomes the thermal entropy in $\hH$. In
section \ref{newstor}, we demonstrate that the same results apply when
we begin with the CFT in a cylindrical background geometry, \ie
$R\times S^{d-1}$.

\subsection{Entanglement entropy in flat space} \label{flat}

Consider an arbitrary quantum field theory  in $d$-dimensional
Minkowski space $R^{1,d-1}$. As shown in figure \ref{pictx}, we
introduce a smooth entangling surface $\Sigma$ which divides the $t=0$
time slice into two parts, a region $V$ and its complement $\bar V$.
Upon integrating out the degrees of freedom in $\bar V$, we are left
with the reduced density matrix $\rho$ describing the remaining degrees
of freedom in the region $V$. The entanglement entropy across $\Sigma$
is then just the von Neumann entropy of $\rho$, \ie
 \be
S_\Sigma=-\textrm{tr}(\rho \,\log\rho)\,.
 \labell{entan7}
 \ee
However, we note that the definition and interpretation of these
objects in the continuum QFT requires a regularization, as will become
apparent below.
\FIGURE[t]{
\includegraphics[width=0.8\textwidth]{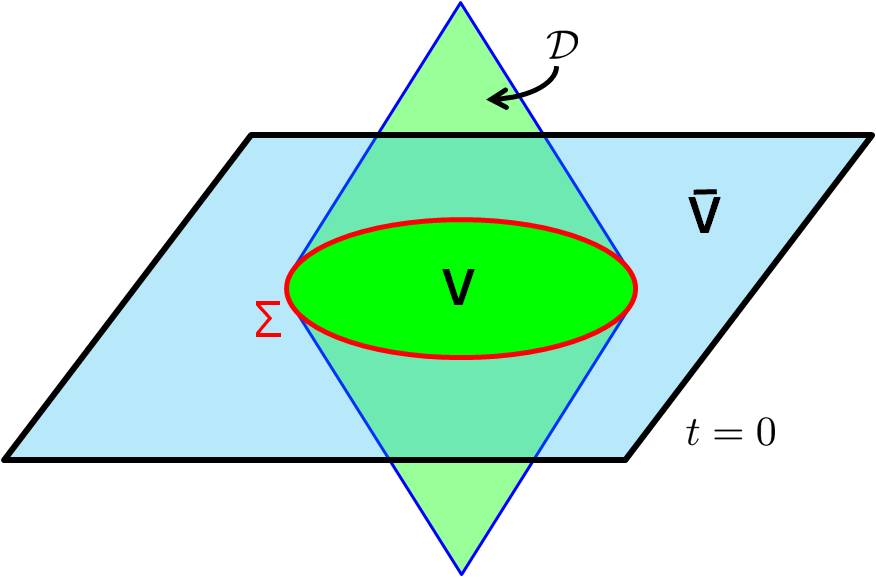}
\caption{The division of the $t=0$ time slice into two regions $V$ and
$\bar V$.} \label{pictx}}

Since the reduced density matrix is both hermitian and positive
semidefinite, it can be expressed as
 \be
\rho=e^{-H} \labell{important}
 \ee
for some hermitian operator $H$. In the literature on axiomatic quantum
field theory, $H$ is known as the modular Hamiltonian
\cite{haag}.\footnote{The same operator, referred to as the
`entanglement Hamiltonian', has recently also appeared in studies of
the `entanglement spectrum' of topological phases of matter
\cite{cmt}.} We emphasize that generically $H$ is not a local operator.
That is, it can not be represented as some local expression constructed
with the fields on $V$. However, the modular Hamiltonian still plays an
important role since the unitary operator $U(s)=\rho^{i s}=e^{-i H s}$
generates a symmetry of the system. One easily sees that
\begin{equation}
\textrm{tr}(\rho\, U(s) {\cal O} U(-s))
=\textrm{tr}(\rho\, {\cal O}) \labell{aaa}\,,
\end{equation}
for any operator ${\cal O}$ localized inside $V$. In fact, because of
causality, this symmetry group transforms the algebra of operators
inside the causal development\footnote{The causal development of $V$,
which we denote $\D$, is the set of all points $p$ for which all causal
curves through $p$ necessarily intersect $V$.}
of $V$ into itself. In the algebraic approach to QFT, this
one-parameter group of transformations $U(s)$ is called the modular
group \cite{haag}. Further if these transformations are extended to
complex parameters, one finds that correlators obey the KMS
(Kubo-Martin-Schwinger) periodicity relation in imaginary time.
Defining ${\cal O}(s)= U(s) {\cal O} U(-s)$, this relation is easily
established with
\begin{equation}
\textrm{tr}(\rho\, {\cal O}_1(i){\cal O}_2)=\textrm{tr}
(\rho\, U(i){\cal O}_1 U(-i) {\cal O}_2)=\textrm{tr}(\rho\,{\cal O}_2
{\cal O}_1) \,.
  \labell{bbb}
\end{equation}
The last equality follows using $U(i)=\rho^{-1}$, $U(-i)=\rho$ and the
cyclicity of the trace. Hence formally we can say the state $\rho$ is
thermal with respect to the time evolution dictated by the internal
symmetry $U(s)$, with an inverse temperature $\beta=1$. However, we
emphasize that these are formal expressions. As noted above,
generically $H$ is not local and $U(s)$ does not generate a local
(geometric) flow on $\D$. For example, if we begin with a local
operator defined at a point, ${\cal O}=\phi(x)$, then generally the
operator $O(s)$ will no longer have this simple form of being defined
at a point.

However, there are special cases where the modular flow and the modular
Hamiltonian are in fact local. One well-known example is given by
Rindler space $\R$, \ie the wedge of Minkwoski space corresponding to
the causal development of the half space $X^1>0$. In this case for {\it
any QFT}, the modular Hamiltonian is just
the boost generator in the $X^1$ direction. This result is commonly
known as the Bisognano-Wichmann theorem \cite{bisognano}. In this case
then, the modular transformations act geometrically in Rindler space.
They map the algebra of operators ${\cal A}(B)$ localized in a region
$B\subseteq {\R}$ to the algebra of operators ${\cal A}(B_{s})$ in the
region $B_{s}$,
\begin{equation}
U(s){\cal A}(B)U(-s)={\cal A}(B_{s})\,,\labell{modularflowx}
\end{equation}
where $B_{s}$ is the mapping of $B$ by the boost transformation. More
explicitly, the modular flow is given by
\begin{equation}
X^\pm(s)=X^\pm e^{\pm 2 \pi s}\,,\labell{modu}
\end{equation}
where $X^{\pm}\equiv X^1\pm X^0$ are the null coordinates with $0\le
X^\pm< +\infty$ in ${\R}$. Of course, the modular flow leaves all other
coordinates invariant, \ie $X^i(s)=X^i$ for $i=2,...,d-1$.

Interpreted in the sense of Unruh \cite{Unruh}, the state in $\R$ is
thermal with respect to the notion of time translations along the boost
orbits. If we choose conventional Rindler coordinates,
\begin{equation}
X^\pm(s)=z \,e^{\pm \tau/R}\,,\labell{modu2}
\end{equation}
the Minkowski space metric becomes
 \be
ds^2=dX^+\,dX^-+\sum_{i=2}^{d-1} \left(dX^i\right)^2
=-\frac{z^2}{R^2}\,d\tau^2+dz^2+\sum_{i=2}^{d-1}\left(dX^i\right)^2\,.
 \labell{rindler}
 \ee
We introduced an arbitrary scale $R$ above in eq.~\reef{modu2} to
ensure that $\tau$ has the standard dimensions of length. With this
choice, the Rindler state is thermal with respect to $H_\tau$, the
Hamiltonian generating $\tau$ translations, with a temperature $T=1/(2
\pi R)$. Hence the density matrix can be simply written as the thermal
density matrix
 \be
\rho_{\ssc \R}=\frac{e^{-\beta H_\tau}}{Z} \qquad {\rm where}\ Z={\rm
tr} \left(e^{-\beta H_\tau}\right)\,.
 \labell{density}
 \ee
With this notation, the modular flow \reef{modu} on $\R$ simply
corresponds to the time translation
 \be
\tau\rightarrow \tau+ 2 \pi R\,s
 \labell{translate0}
 \ee
and the modular Hamiltonian $H_{\ssc \R}$ is simply related to $H_\tau$
with $H_{\ssc \R}=2\pi R\,H_\tau\,+\,\log Z$.

In the following, we are particularly interested in the entanglement
entropy for the case where the entangling surface $\Sigma$ is a
($d$--2)-dimensional sphere of radius $R$. Hence the region $V$ becomes
the ball bounded by this $S^{d-2}$. Let us define the null coordinates
$x^\pm\equiv r\pm t$ with $t=x^0$ and the radial coordinate
$r=\sqrt{(x^1)^2+...+(x^{(d-1)})^2}$. Then the causal development $\D$
of the ball is the spacetime region defined by $\lbrace x^+\le
R\rbrace\, \cap\,\lbrace x^-\le R\rbrace$ --- implicitly, we are
assuming $x^++x^-=2r\ge0$ in this definition. Further we wish to
consider the special case where the entanglement entropy is calculated
in this geometry for a conformal field theory. In this case, the
modular Hamiltonian on $\D$ will in fact be a local operator.

This last fact can be derived making use of the previous result for the
Rindler wedge $\R$. To begin, we observe that there is a special
conformal transformation (and translation) which maps the Rindler wedge
to the causal development $\D$ of the ball (\eg see \cite{haag,Hislop})
\begin{equation}
x^\mu=\frac{X^\mu-(X\cdot X) C^\mu}{1-2 (X\cdot C)+(X\cdot X)(C\cdot
C)}+2 R^2 C^\mu\,,\labell{doce}
\end{equation}
with $C^\mu=(0,1/(2R),0,...0)$. It is straightforward to show that
$X^\pm\ge0$ covers  $x^\pm\le R$. Explicitly, eq.~\reef{doce} yields
$ds^2=\eta_{\mu\nu}dX^\mu dX^\nu=\Omega^2\, \eta_{\mu\nu}dx^\mu dx^\nu$
where the conformal pre-factor can be written as
 \bea
\Omega&=&1-2 (X\cdot C)+(X\cdot X)(C\cdot C)\,.
\nonumber\\
&=&\ \left(1+2(x\cdot C)+(x\cdot x)(C\cdot C)\right)^{-1}\,.
 \labell{Omega}
 \eea
Further eq.~\reef{doce} maps the flow \reef{modu} to the following
geometric flow in ${\cal D}$,
\begin{equation}
x^\pm(s)=R\,\frac{ (R+x^\pm)-e^{\mp 2\pi s} (R-x^\pm)}{
 (R+x^\pm)+e^{\mp 2\pi s} (R-x^\pm)} \,. \labell{fl}
\end{equation}

Now it is not difficult to show that the induced flow (\ref{fl}) gives
the modular flow of the CFT on ${\cal D}$. Recall that there is a
unitary operator, which we denote $U_{0}$, in the CFT which implements
the conformal transformation associated eq.~(\ref{doce}). For example,
the primary operators of the CFT transform locally as (considering
spinless operators for simplicity)
 \begin{equation}
\phi(x)=\Omega(X)^{\Delta}\, U_{0}\,\phi(X)\,U_{0}^{-1}
 \labell{transform}
 \end{equation}
where $\Delta$ is the scaling dimension of the field. Since this
mapping is a conformal symmetry of Minkowski space, it leaves the
vacuum state invariant in a conformal theory, \ie $U_{0}\vert
0\rangle=\vert 0 \rangle$. At a more practical level, vacuum
correlators on $\R$ are mapped to vacuum correlators on $\D$
\be
\langle \phi_1(x_1)\cdots\phi_n(x_n)\rangle
=\Omega(X_1)^{\Delta_1}\cdots\Omega(X_n)^{\Delta_n}\, \langle
\phi_1(X_1)\cdots\phi_n(X_n)\rangle \,.
 \labell{transi}
 \ee
Now we
may apply $U_0$ to construct the quantum operator generating the
modular flow on ${\cal D}$ with
\begin{equation}
U_{\ssc \D}(s)=U_{0}\,U_{\ssc \R}(s)\,U_{0}^{-1}\,.\labell{ff0}
\end{equation}
One can show that $U_{\ssc \D}(s)$ generates the `modular
transformations' for the sphere $\Sigma$ \cite{Hislop}. To confirm that
eq.~\reef{ff0} yields the modular operator, we must verify that two
conditions are satisfied \cite{haag}: {\it i}) the transformation must
be a symmetry of the correlators, as in eq.~\reef{aaa}, and {\it ii})
the correlators must obey the KMS periodicity, as in eq.~\reef{bbb}.
The first condition eq.~\reef{aaa} is evident, since (\ref{ff0}) is
just a particular conformal symmetry of the theory which keeps the
sphere invariant.

Turning to the second condition, we note that the
correlator on the right hand side of (\ref{transi}) satisfies the KMS condition with
respect to the flow (\ref{modu}). Further combining eqs.~\reef{modu}
and \reef{Omega}, it is also clear that each of the pre-factors
involving $\Omega$ is also periodic in $s$ with imaginary period $i$.
Hence, the correlator on the left hand side also satisfies the KMS
condition but with respect to the induced flow (\ref{fl}). Hence given
that conditions ({\it i}) and ({\it ii}) are both satisfied, we
conclude that (\ref{ff0}) are indeed the modular transformations and the geometric flow \reef{fl} is the modular flow
on $\D$.

 Acting on a (spinless) primary operator in the
CFT, one finds
 \begin{equation}
U_{\ssc \D}(s)\,\phi(x[s_0])\,U_{\ssc \D}(-s)=\Omega(x[s_0])^{\Delta}\,
\Omega(x[s_0+ s])^{-\Delta}\, \phi(x[s_0+ s])\,,
 \labell{transform2}
 \end{equation}
using eq.~\reef{transform}. Here, our notation $x[s_0]$ indicates that
the position of the operator flows according to eq.~\reef{fl}. Note
then that $U_{\ssc \D}(s)$ acts both to translate the operator along
the geometric flow \reef{fl} and multiplies it with a particular
(c-number) pre-factor. The key point, however, is that $U_{\ssc \D}(s)$
remains a local transformation, taking an operator defined at a point
to that same operator defined at a new point.

In fact, we can produce an explicit expression for the modular
Hamiltonian at this point. The construction is simplest if we focus on
$V$, the ball of radius $R$, in the time slice $x^0=0$. Since the
modular Hamiltonian is the generator for the transformation
\reef{transform2}, it will produce an infinitesimal shift $\delta s$ on
the surface $x^0=0$:
\begin{equation}
\delta x^0= 2 \pi \frac{R^2-r^2}{2 R} \delta s\qquad
{\rm and}\qquad \delta r = 0\,. \labell{shift9}
\end{equation}
At the same time, an infinitesimal shift $\delta s$ produces a
pre-factor in eq.~\reef{transform2} which takes the form
 \be
\left.\Omega(x[s_0])^{\Delta}\, \Omega(x[s_0+\delta
s])^{-\Delta}\right|_{x^0=0} \simeq 1-\Delta\left.
\frac{\partial_s\Omega}{\Omega}\right|_{x^0=0} \delta s\,.
 \labell{prefact}
 \ee
However, using eqs.~\reef{Omega} and \reef{shift9}, it is
straightforward to show that ${\partial_s\Omega}=0$ when evaluated on the
$x^0=0$ surface and so the pre-factor is simply 1 at this order. Hence
the modular Hamiltonian simply induces the infinitesimal flow
\reef{shift9} away from $x^0=0$ and we can identify the corresponding
operator in the CFT as
\begin{equation}
H_{\ssc \D}=2 \pi \int_V d^{d-1}x \frac{(R^2-r^2)}{2R} T^{00}(x) + c'
\,,\labell{hmodular}
\end{equation}
where $T^{\mu\nu}$ is the conformal {\it traceless} stress tensor and
$c'$ is some constant added to ensure that the corresponding density
matrix is normalized with unit trace. Hence this expression explicitly
shows the modular Hamiltonian as a local operator, for a CFT in the
causal development $\D$.

One point that we feel is worth emphasizing is that the standard CFT
correlators in $\D$ transform `covariantly' under $U_{\ssc \D}(s)$
--- that is, they are {\it not} invariant, as one might na\"ively surmise
from eq.~\reef{aaa}. This is simply the observation that if we begin
with $\cO= \phi_1(x_1[s_0]) \cdots \phi_n(x_n[s_0])$ in eq.~\reef{aaa},
then eq.~\reef{transform2} dictates that generally $\cO(s)\not=
\phi_1(x_1[s_0+s]) \cdots \phi_n(x_n[s_0+s])$. Instead, the correlators
of (spinless) primary operators transform as
 \bea
&&\Omega(x_1[s_0])^{-\Delta_1}\cdots\Omega(x_n[s_0])^{-\Delta_n}\,\langle
\phi_1(x_1[s_0])\cdots\phi_n(x_n[s_0])\rangle
 \labell{trans}\\
&&\qquad=\Omega(x_1[s_0+s])^{-\Delta_1}\cdots\Omega(x_n[s_0+s])^{-\Delta_n}\,
\langle \phi_1(x_1[s_0+s])\cdots\phi_n(x_n[s_0+s])\rangle\,.
 \nonumber
 \eea

Clearly, the idea of `transplanting' the modular flow with a conformal
transformation can be used to obtain the modular Hamiltonian (and the
density matrix) for a CFT in other conformally connected geometries.
Note that the state in the transformed space has to be chosen as the
conformally transformed state, which is the one that makes
(\ref{trans}) work. In the present case of a mapping between the
Rindler wedge $\R$ and the causal development ${\cal D}$, the
transformation is a conformal symmetry of Minkowski space and the
transformed state is again the Minkowski vacuum. Of course, for
operators inside ${\cal D}$, the vacuum coincides with the density
matrix $\rho$, meaning that they both give the same expectation values
$\langle0\vert {\cal O}\vert 0 \rangle=\textrm{tr}(\rho\, {\cal O})$.

\subsection{Thermal behaviour in $R\times H^{d-1}$} \label{therm}

Next we would like to extend this approach of transplanting modular
flows to relate the density matrix of a CFT on $\D$ to that on a new
geometry $R\times H^{d-1}$, which we will denote as $\hH$ in the
following. In particular, we will show that beginning with the
Minkowski vacuum for an arbitrary CFT, the density matrix on $\D$
becomes a thermal density matrix on $\hH$. In fact, our result is a
generalization of previous observations made in ref.~\cite{candow}.
There the generation of a thermal state by conformal mappings was
observed for free conformal field theories in $d=4$.

One indication that the claim above holds comes from first returning to
the Rindler wedge \reef{rindler}. Here, we can write the metric as
 \be
ds^2=\Omega^2\left(-d\tau^2+\frac{R^2}{z^2}\left[dz^2+
\sum_{i=2}^{d-1}\left(dX^i\right)^2
\right]\right)\,,
 \labell{rindler2}
 \ee
where $\Omega=z/R$. Hence with a conformal transformation which
eliminates the pre-factor $\Omega^2$, the Rindler metric is mapped
precisely to the metric on $R\times H^{d-1}$. Note that the scale $R$
now sets the curvature scale of the hyperbolic plane $H^{d-1}$. As in
the previous section, there is a unitary operator which maps the CFT
from $\R$ to $\hH$, which we will denote $U_{1}$. Further we may again
apply $U_1$ to construct the operator generating the modular flow on
${\hH}$ with
\begin{equation}
U_{\ssc \hH}(s)=U_{1}\,U_{\ssc \R}(s)\,U_{1}^{-1}\,.\label{ff2}
\end{equation}
In this particular case, the conformal mapping is time (\ie $\tau$)
independent and so $U_1$ acts `trivially' on the modular Hamiltonian.
That is, $H_{\ssc \hH}=2\pi R\,H_\tau\,+\,\log Z$ where $H_\tau$ is now
the generator of $\tau$-translations in $\hH$. Therefore the new
density matrix $\rho_{\ssc \hH}$ inherits the same thermal character as
in eq.~\reef{density}. Hence combining this result with those in the
previous section, we can map the reduced density matrix on $\D$ to a
thermal density matrix on $\R$ and then to a thermal density matrix on
$\hH$. However, let us step back and establish the relationship between
$\rho_{\ssc\D}$ and $\rho_{\ssc\hH}$ directly.

First, we present the conformal transformation which maps $\D$ to
$\hH$. We start with the flat space metric in polar coordinates:
 \beq
ds^2=-dt^2+dr^2+r^2\,d\Omega^2_{d-2}\,,
 \labell{flat0}
 \eeq
where $d\Omega^2_{d-2}$ is the line element on a round $S^{d-2}$ with
unit curvature. Our entangling surface $\Sigma$ is again the sphere
$r=R$ on the surface $t=0$. Now we make the coordinate transformation
 \beqa
t&=&R\,\frac{\sinh(\tau/R)}{\cosh u+\cosh(\tau/R)}\,,
 \labell{trn1}\\
r&=&R\,\frac{\sinh u}{\cosh u+\cosh(\tau/R)}\,.
 \nonumber
 \eeqa
One can readily verify the above metric \reef{flat0} becomes
 \beqa
ds^2&=&\Omega^2\,\left[-d\tau^2+ R^2\,\left(du^2+\sinh^2\!
u\,d\Omega^2_{d-2}\right)\right]
 \nonumber\\
{\rm where}&&\ \ \Omega=(\cosh u+\cosh(\tau/R))^{-1}\,.
 \labell{hyper0}
 \eeqa
After the conformal transformation which eliminates the pre-factor
$\Omega^2$, we again recognize the resulting line element as the metric
on $R\times H^{d-1}$. The curvature on the latter hyperbolic space is
 \beq
R_{ij}{}^{k\ell}=-\frac{1}{R^2}\left(\delta_i{}^k\delta_j{}^\ell
-\delta_i{}^\ell\delta_j{}^k\right)\,.
 \labell{curve9}
 \eeq
We note that fixing the curvature scale to match $R$, the radius of the
sphere $\Sigma$, is an arbitrary but convenient choice.
Finally we observe that
 \beqa
\tau\rightarrow\pm\infty\,:&&\quad(t,r)\rightarrow(\pm R,0)\nonumber\\
u \rightarrow\infty\,:&&\quad(t,r)\rightarrow(0,R)\nonumber
 \eeqa
Hence the new coordinates cover precisely $\D$, the causal development
of the region inside of $\Sigma$. Hence we have our conformal mapping
from $\D$ to $\hH$. We are again considering an arbitrary CFT and so
there is a unitary transformation $U_2$ implementing the conformal
transformation (\ref{trn1}) on the Hilbert space of the
CFT.\footnote{Of course, this operator is related to those appearing in
the previous discussion by $U_2=U_1U_0^{-1}$.}

Now, we wish to relate the two reduced density matrices: $\rho_{\ssc
D}$ describing the vacuum state on $\D$ and $\rho_{\ssc \hH}$ for the
corresponding state on $\hH$. In particular, we wish to establish that
the latter corresponds to a thermal density matrix with temperature
$T=1/(2\pi R)$. For the latter to hold, we must verify two conditions:
First, the modular flow in $\hH$ must correspond to ordinary time
translations and these translations must be a symmetry of the
correlators. Second, the correlators on $\hH$ must be periodic for an
imaginary shift of $\tau$ by $2\pi R$.

As a first step, we use eq.~\reef{trn1} to write
 \be
x^\pm = R\frac{1-e^{-v^\pm}}{1+e^{-v^\pm}}
 \labell{trn1x}
 \ee
where we have defined $v^\pm\equiv u\pm (\tau/R)$ --- recall from above
that $x^\pm=r\pm t$. From these expressions, it is straightforward to
show that the modular flow \reef{fl} on $\D$ corresponds to a time
translation
 \be
\tau\rightarrow \tau+ 2 \pi R\,s
 \labell{translate}
 \ee
in $\hH$ --- just as in eq.~\reef{translate0}. Hence as desired, the
modular Hamiltonian induces a flow along time translations. However,
for a thermal state, this flow must correspond to a symmetry of the
correlators, \ie it is not enough that the correlators transform
covariantly under the modular flow. The primary fields transform by the
analog of eq.~\reef{transform} replacing $U_0$ by $U_2$ and hence we
can write the relation for the correlation functions under the
conformal transformation as
 \begin{equation}
\langle \phi^\prime_1(x_1^\prime)\cdots\phi^\prime_n(x_n^\prime)\rangle_\hH =
\Omega(x_1^\prime)^{\Delta_1}\cdots\Omega(x_n^\prime)^{\Delta_n}\,
\langle \phi_1(x_1)\cdots\phi_n(x_n)\rangle_\D\,.
 \labell{transt}
\end{equation}
where $\Omega(x^\prime)$ is the conformal pre-factor in
eq.~(\ref{hyper0})). Now the factors $\Omega$ in (\ref{transt}) are not
invariant under shifts of $\tau$. However, we also noted above in
eq.~\reef{trans} that the modular flow on $\D$ transforms the
correlators on the right-hand side non-trivially. Having the connection
\reef{translate} between translations in $\tau$ and $s$, a simple
calculation shows the $\tau$ dependence of the pre-factors precisely
cancels that of the correlators on $\D$. Hence the conformally mapped
correlators on $\hH$ are in fact invariant under shifts in $\tau$.
Hence the modular flow in $\hH$ is just the ordinary time translations
\reef{translate}. Examining eq.~\reef{transt} further, we find the
second requirement, that the correlators on $\hH$ must be periodic for
an imaginary shift of $\tau$ by $2\pi R$ also follows. First, given
eq.~\reef{hyper0}, it is clear that each of conformal pre-factors
satisfies this condition. Next the correlators on $\D$ satisfy the KMS
condition \reef{bbb} and from eq.~\reef{translate}, it follows that the
imaginary shift of $s$ by $i$ corresponds to the desired shift of
$\tau$.

Hence the two requirements above are both satisfied and so we have
established the desired result: The conformal mapping \reef{hyper0}
takes the CFT in the Minkowski vacuum on $\D$ to a thermal state on
$\hH$ where the latter is thermal with respect to the standard
Hamiltonian $H_\tau$ with a physical temperature $T=1/(2 \pi R)$. That
is, we have $\rho_{\ssc \hH}=e^{-\beta H_\tau}/Z$. Now the unitary
operator $U_2$ will map between the reduced density matrix on $\D$ and
the thermal density matrix on $\hH$. More explicitly, we may write the
density matrix on $\D$ as
 \be
\rho_{\ssc\D}\equiv e^{-H_{\ssc\D}}=\frac1Z \,U^{-1}_2e^{-\beta H_\tau}
U_2\,.
 \labell{newrho}
 \ee
Since the von Newman entropy is invariant under unitary
transformations, the entanglement entropy across the sphere $\Sigma$ in
flat space is then equal to the corresponding thermal entropy of the
Gibbs state in $\hH$.

However, some care must be exercised for the equality of the entropies
to hold since both of these quantities are divergent. In particular, in
order to maintain the equality, we must also use the conformal
transformation to map between the cut-off procedures in two spaces. In
the case of entanglement entropy \reef{entan7}, there is a UV
divergence at the boundary $\Sigma$ and so we need to introduce a short
distance cut-off scale $\delta$. That is, we only take contributions
down to $r=R-\delta$ where $\delta/R\ll 1$.\footnote{A regularization
of the entanglement entropy implementing a distance cutoff to the
entangling sphere can be made precise using the mutual information
\cite{ch1}.} In the case of the thermal entropy, there is an IR
divergence because we have a uniform entropy density but the volume of
the spatial slices, \ie $ H^{d-1}$, is infinite. Hence we regulate the
entropy by integrating out to some maximum radius, $u=u_{max}$ where
$u_{max}\gg 1$. Now, consistency demands that the cut-off's should be
related by the conformal mapping between the two spaces. If we focus on
the $t=0$ slice (or equivalently the $\tau=0$ slice), then
eq.~\reef{trn1} yields
 \beq
 R-\delta=R\,\frac{\sinh u_{max}}{\cosh u_{max}+1}\,.
 \labell{cutoff1}
 \eeq
After a bit of algebra, we find
 \beq
 \exp(-u_{max})=\frac{\delta/R}{2-\delta/R}\simeq\frac{\delta}{2R}\,.
 \labell{cutoff2}
 \eeq
It is interesting that here the conformal mapping has introduced a
UV/IR relation between the relevant cut-offs in the two CFT states. Of
course, similar relations commonly occur in the AdS/CFT correspondence
but here we have a purely CFT calculation.

\subsection{Entanglement entropy in a cylindrical background} \label{newstor}

In this section, we develop the analogous account of the
`thermalization by conformal mapping' found above for the case where
the original background geometry is $R\times S^{d-1}$. That is, we wish
to calculate the entanglement entropy across a sphere of a fixed
angular size embedded in the $t=0$ slice of the static Einstein
universe. As before, we map the causal development of the interior of
this sphere to $R\times H^{d-1}$ with a conformal transformation and
show that the resulting density matrix describes a thermal state. Hence
the entanglement entropy becomes the thermal entropy of the Gibbs state
in $\hH$. Below we only present the salient features of the proof as
conceptually it is completely analogous to the discussion above.

We start with the metric in $d$-dimensional cylindrical spacetime with
topology $R\times S^{d-1}$:
 \beq
ds^2=-dt^2+R^2\,\left(d\theta^2+\sin^2\!\theta\,d\Omega^2_{d-2}\right)\,.
 \labell{cyl0}
 \eeq
The entangling surface $\Sigma$ will now be the sphere
$\theta=\theta_0$ on the surface $t=0$. Now make the coordinate
transformation
 \beqa
\tan(t/R)&=&\frac{\sin\theta_0\,\sinh(\tau/R)}{\cosh
u+\cos\theta_0\,\cosh(\tau/R)}\,,
 \labell{trn3}\\
\tan\theta&=&\frac{\sin\theta_0\,\sinh u}{\cos\theta_0\,\cosh u +\cosh
(\tau/R)}\,.
 \nonumber
 \eeqa
One can readily verify the above metric \reef{cyl0} becomes
 \beqa
ds^2&=&\Omega^2\, \left[-d\tau^2+ R^2\,\left(du^2+\sinh^2\!
u\,d\Omega^2_{d-2}\right)\right]
 \labell{hyper2}\\
{\rm where}&& \Omega^2=\frac{\sin\theta_0^2}{\left(\cosh
u+\cos\theta_0\,\cosh(\tau/R)\right)^2 + \sin\theta_0^2\,\sinh^2\!u}\,.
 \nonumber
 \eeqa
Hence, after eliminating the conformal factor $\Omega^2$ in the first
line, we again recognize the final line element as the metric on
$R\times H^{d-1}$,
 \beq
 ds'^2=-d\tau^2+ R^2\,\left(du^2+\sinh^2\! u\,d\Omega^2_{d-2}\right)\,,
 \labell{final}
 \eeq
precisely as in eq.~\reef{hyper0}. The curvature scale of the
hyperbolic space is $R$, as in eq.~\reef{curve9}. As before, we note
that this is an arbitrary but convenient choice for the curvature. In
this case, $R$ also corresponds to the radius of curvature of the
$S^{d-1}$ in the Einstein universe, rather than the size of the
entangling sphere $\Sigma$, as we will see below.

Examining what portion of the original Einstein universe \reef{cyl0} is
covered by the new coordinates, we find
 \beqa
\tau\rightarrow\pm\infty\,:&&\quad(t,\theta)\rightarrow
(\pm R\theta_0,0)\nonumber\\
u \rightarrow\infty\,:&&\quad(t,\theta)\rightarrow
(0,\theta_0)\nonumber
 \eeqa
Hence, the new coordinates cover precisely the causal development $\D$
of the ball enclosed by the entangling sphere $\Sigma$. As in section
\ref{newstor}, examining the effect of the conformal mapping on CFT
correlators, we find that the vacuum correlators in $\D$ become thermal
correlators on $\hH$ with temperature $T=1/(2\pi R)$.

As before, the entanglement entropy across the sphere $\Sigma$ in the
Einstein space matches the thermal entropy of the Gibbs state in $\hH$
but we must be careful in regulating the two expressions. In the case
of entanglement entropy, there is a UV divergence at the surface
$\Sigma$ and we need to introduce a short distance cut-off. The natural
cut-off is a minimal angular size $\delta\theta\,(\ll 1)$ and only
taking contributions out to $\theta=\theta_0-\delta\theta$. As before,
the thermal entropy has an IR divergence because we have a uniform
entropy density in the the infinite volume of the spatial $H^{d-1}$
slices. We again regulate this entropy by only integrating out to
$u=u_{max}$ where $u_{max}\gg 1$. Now consistency demands that the
cut-off's should be related by the conformal mapping between $\D$ and
$\hH$. If we focus on the $t=0$ slice (or equivalently the $\tau=0$
slice), then eq.~\reef{trn3} yields
 \beq
 \tan(\theta_0-\delta\theta)=\frac{\sin\theta_0\,\sinh u_{max}}{\cos\theta_0\,
 \cosh u_{max}+1}\,.
 \labell{cutoff3}
 \eeq
With a bit of algebra, we find
 \beq
 \exp(-u_{max})=\frac{1}{2\sin\theta_0}
 \frac{\tan\delta\theta+\tan\theta_0(\sec\delta
 \theta-1)}{1-\cot2\theta_0\,\tan\delta\theta}
 \simeq\frac{\delta\theta}{2\sin\theta_0}\,.
 \labell{cutoff4}
 \eeq
Hence as in the previous section, we have an interesting UV/IR relation
between the cut-offs in the two CFT calculations.

\section{The AdS story} \label{ads}

In the previous section, we related the problem of calculating
entanglement entropy for a spherical entangling surface to calculating
the thermal entropy of a Gibbs state in $R\times H^{d-1}$, where the
temperature is of the same order as the curvature scale of the
hyperbolic geometry. While this is quite general result that applies
for any arbitrary CFT, it seems that we have only replaced one
difficult problem with another equally difficult problem. However, we
are particularly interested in applying this result to the AdS/CFT
correspondence. In this framework, the standard holographic dictionary
suggests that the thermal state in the boundary CFT is dual to a black
hole in the bulk gravity theory. Hence if we are able to identify the
corresponding black hole, the thermal entropy of the CFT can be
calculated as the horizon entropy of the black hole.

Precisely this problem was encountered in \cite{cthem,cthem2} in a
related calculation of the entanglement entropy for a specific
geometry. In this case, the boundary CFT was placed in a cylindrical
space, $R\times S^{d-1}$
--- that is, the (complete) boundary of AdS$_{d+1}$. The problem was
then to determine the entanglement entropy when the
($d$--1)-dimensional sphere was divided in half along the equator, \ie
the entangling surface was chosen as a maximal $S^{d-2}$ in a constant
time slice. In \cite{cthem,cthem2}, it was argued that the entanglement
entropy could be identified with the horizon entropy of a topological
AdS black hole, for which the large radius limit (at fixed time)
covered precisely one half of the boundary $S^{d-1}$. In fact, the
topological black hole was simply an $R\times H^{d-1}$ foliation of the
AdS$_{d+1}$ geometry.

Given the CFT discussion of section \ref{cft}, we now have a clearer
understanding of the calculations in \cite{cthem,cthem2} and in turn,
the calculations there suggest the necessary approach to implement our
results from the previous section in a holographic setting. In
particular, we begin with the boundary CFT in its vacuum either in
Minkowski space $R^{1,d-1}$ or the cylindrical space $R\times S^{d-1}$.
The dual gravity description is simply the pure AdS$_{d+1}$ geometry in
a coordinate system which foliates the spacetime with $R^{1,d-1}$ or
$R\times S^{d-1}$ surfaces. To calculate the entanglement entropy of
the CFT across some sphere $\Sigma$ in the boundary geometry, we must
next find a new foliation of the AdS$_{d+1}$ in terms of $R\times
H^{d-1}$. This foliation is chosen to cover the ball enclosed by
$\Sigma$ in the asymptotic boundary geometry. In fact, the causal
development $\D$ of this ball will be covered in the asymptotic limit
of the $R\times H^{d-1}$ foliation. Implicitly, we will have
implemented the conformal mapping of the causal development $\D$ of the
ball to $R\times H^{d-1}$ in the boundary CFT with the bulk coordinate
transformation between the two foliations of the AdS space. As noted in
\cite{cthem,cthem2}, the new hyperbolic foliation can be interpreted in
terms of a topological black hole \cite{roberto,topbh} and so on the
$R\times H^{d-1}$ background, the boundary CFT is naturally seen to be
in a thermal state. The temperature of the latter thermal state is
given by the Hawking temperature of the black hole horizon and we will
find that we precisely reproduce the result of section \ref{cft}:
$T=1/(2\pi R)$. Further, since as discussed in section \ref{cft}, the
entanglement entropy across the sphere $\Sigma$ can be calculated as
the thermal entropy of the Gibbs state in the $R\times H^{d-1}$
geometry, the same entropy is given by the horizon entropy of the bulk
black hole in the holographic setting.

Let us emphasize that this approach gives us a derivation of the
entanglement entropy of the holographic CFT, in the case of a spherical
entangling surface. Given the insights of the previous section, we are
simply applying the standard AdS/CFT dictionary to implement the
results there in a holographic framework. Further let us note that if
the bulk theory was Einstein gravity, then our results of the
entanglement entropy for this particular set of geometries would
precisely match the holographic entanglement entropy calculated with
the extremal area prescription \reef{define} conjectured by Ryu and
Takayanagi \cite{ryut0,ryut1}. In this regard, our results provide a
nontrivial confirmation of their proposal.

In fact, since we have a derivation of the entanglement entropy, we do
not need to limit our discussion to Einstein gravity. Hence in the
following discussion, we allow the bulk gravity to be described by any
arbitrary covariant action of the form
 \be
I=\int d^{d+1} x \sqrt{-g}\, {\cal L}(g^{ab},R^{ab}{}_{cd}, \nabla_e
R^{ab}{}_{cd}, \cdots, matter)\,.
 \labell{covariant}
 \ee
Hence one might apply our analysis in the context of the low energy
effective action in string theory, where the contributions of higher
curvature terms are controlled to make small corrections to the results
of the Einstein theory \cite{stringx}. Alternatively, our results would
also be applicable to a situation where higher curvature contributions
are finite, as in the recent studies of the AdS/CFT correspondence
\cite{holo} with Lovelock \cite{lovel} or quasi-topological \cite{old1}
gravity. In the following, we will assume that the couplings of the
above theory are chosen so that the theory has a vacuum solution which
is AdS$_{d+1}$ spacetime with a curvature scale $L$. Implicitly, we
will also assume that the bulk couplings are further constrained so
that the boundary CFT has physically reasonable properties (\eg it
should be causal and unitary) --- for example, see
\cite{cthem2,old1,holo}.

An essential step in the following will be calculating the horizon
entropy of the bulk black hole. In general, the horizon entropy can be
calculated using Wald's entropy formula \cite{WaldEnt}
 \beq
S = -2 \pi \int_\mt{horizon}\! d^{d-1}x\sqrt{h}\
\frac{\partial{\mathcal{L}}}{\partial R^{a b}{}_{c d}}\,\hat{\veps}^{a
b}\,\hat{\veps}_{c d}\,,
 \labell{Waldformula}
 \eeq
which can be applied for any (covariant) action, as assumed above in
eq.~\reef{covariant}. Note that $\hat{\veps}_{a b}$ denotes the
binormal to the horizon. Of course, as described above, the case of
interest is a topological black hole which in fact corresponds to a
$R\times H^{d-1}$ foliation of AdS$_{d+1}$. In this case, the integrand
in eq.~\reef{Waldformula} is constant across the horizon and so the
total entropy diverges. Of course, this divergence is expected given
the discussion towards the end of section \ref{therm} and we will
return to this point in the following. However, at this point, we use
some results from \cite{cthem2} to re-express the integrand as
 \be
\left.\frac{\delta \cal L}{\delta R^{ab}{}_{cd}}\,\hat{\veps}^{a
b}\,\hat{\veps}_{c d}\right|_\mt{AdS}=
-\frac{\Gamma(d/2)}{\pi^{d/2}}\,\frac{\ads}{L^{d-1}}
 \labell{adseom2}
 \ee
where $L$ is the AdS curvature scale. The (dimensionless) constant
$\ads$ is a central charge that characterizes the number of degrees of
freedom in the boundary CFT \cite{cthem2}. In the case where $d$, the
dimension of boundary theory, is even, $\ads$ is precisely equal to
$A$, the coefficient of the $A$-type trace anomaly \reef{traza} in the
CFT \cite{cthem2}. We stress that eq.~\reef{adseom2} relies on the fact
that the background geometry in which we are evaluating this expression
is simply the AdS$_{d+1}$ spacetime. Substituting this result into
eq.~\reef{Waldformula} leaves us with
 \beq
S = \frac{2\,\Gamma(d/2)}{\pi^{d/2-1}}\,\frac{\ads}{L^{d-1}}\,
\int_\mt{horizon}\! d^{d-1}x\sqrt{h}\ .
 \labell{Waldformula2}
 \eeq

We now turn to the detailed determination of the hyperbolic foliations
discussed above. We begin by examining the entangling surface being a
sphere in flat space in the next section and then consider the case
with a cylindrical background in section \ref{static}.

\subsection{Entanglement entropy in flat space} \label{flat2}

Here we want to present a holographic calculation that implements the
discussion of entanglement entropy of a CFT in flat space given in
section \ref{flat}. So let us begin with a standard description of the
AdS$_{d+1}$ geometry as the following surface
 \be
-y_{-1}^2-y_0^2+y_1^2+\cdots+y_d^2=-L^2
 \labell{hyperbola}
 \ee
embedded in $R^{2,d}$ with
 \be
 ds^2=-dy_{-1}^2-dy_0^2+dy_1^2+ \cdots+dy_d^2\,.
 \labell{newmet2}
 \ee
Now we can connect this geometry to the standard Poincar\'e coordinates
on AdS$_{d+1}$ space with
 \be
y_{-1}+y_d=\frac{L^2}{z}\,,\qquad y^a=\frac{L}{z}\,x^a\ \ \ {\rm with}\
a=0,\cdots,d-1\,.
 \labell{poincare}
 \ee
Now eq.~\reef{hyperbola} becomes a constraint which yields
 \be
 y_{-1}-y_d=z+\frac{1}{z}\eta_{ab}\,x^a x^b\,.
 \labell{poincare2}
 \ee
The induced metric on the hyperboloid \reef{hyperbola} then becomes the
standard AdS$_{d+1}$ metric
 \be
 ds^2=\frac{L^2}{z^2}\left(dz^2+\eta_{ab}\,dx^a dx^b\right)
 \labell{poincare3}
 \ee
where the bulk spacetime is foliated with slices corresponding to flat
$d$-dimensional Minkowski space. As usual in the AdS/CFT framework, we
take the asymptotic limit $z\to0$ and remove a factor of $L^2/z^2$ from
the boundary metric. This yields $ds^2_\mt{CFT}=\eta_{ab}\,dx^a dx^b$
as the metric in which the dual CFT lives when making our holographic
calculations.

Another useful foliation of AdS$_{d+1}$ is given by
 \beqa
y_{-1}&=&\rho\cosh u\,,\quad y_0=\tr\sinh(\ttau/L)\,,\quad
y_d=\tr\cosh(\ttau/L)\,,\labell{hyperx0}\\
y_1&=&\rho\sinh u\cos\phi_1\,,\ y_2=\rho\sinh u\sin\phi_1\cos\phi_2\,,
\cdots\ y_{d-1}=\rho\sinh u\sin\phi_1\sin\phi_2\cdots\sin\phi_{d-2}\,.
 \nonumber
 \eeqa
In this case, the constraint imposed by eq.~\reef{hyperbola} yields
$\tr^2=\rho^2-L^2$ and the induced metric on the hyperboloid
\reef{hyperbola} becomes
 \be
 ds^2=\frac{d\rho^2}{\frac{\rho^2}{L^2}-1}-\left(\frac{\rho^2}{L^2}-1\right)d\ttau^2
+\rho^2\left(du^2+\sinh^2u\,d\Omega^2_{d-2}\right)
 \labell{hyperx3}
 \ee
where
$d\Omega^2_{d-2}=d\phi_1^2+\sin^2\phi_1\left(d\phi_2^2+\sin^2\phi_2\left(
d\phi_3^2+\cdots\right)\right)$ is the line element on a unit
$S^{d-2}$. Of course, the bracketed expression in eq.~\reef{hyperx3},
which is multiplied by $\rho^2$, corresponds to the line element on a
($d$--1)-dimensional hyperbolic plane $H^{d-1}$ with unit curvature.
Hence as desired, we are foliating the AdS$_{d+1}$ space with surfaces
with a $R\times H^{d-1}$ geometry in the above metric \reef{hyperx3}.
Taking the asymptotic limit and removing a factor of $\rho^2/L^2$, we
find the boundary metric is precisely that given in eq.~\reef{final}
with $R=L$ and $\ttau=\tau$.

For our purposes, the essential feature of the second metric
\reef{hyperx3} is that it can be interpreted as a topological black
hole \cite{roberto,topbh} where the horizon at $\rho=L$ has uniform
negative curvature. We would like to see what portion of the asymptotic
AdS boundary in the Poincar\'e coordinates is covered by the exterior
of this topological black hole. We begin by considering the bifurcation
surface (\ie the intersection of the past and future horizons) in
eq.~\reef{hyperx3} which corresponds to $\rho=L$ and any finite value
of $\ttau$. Combining eqs.~\reef{poincare} and \reef{hyperx0}, we find
on this surface
 \be
 y_{-1}+y_d=\frac{L^2}{z}=L\cosh u\,, \qquad y_1^2+\cdots+y_{d-1}^2=
 \frac{L^2}{z^2}r^2=L^2\sinh^2 u\,.\,
 \labell{connect0}
 \ee
where we have introduced a radial coordinate for the boundary
coordinates, \ie $r^2=(x^1)^2+...+(x^{d-1})^2$. Hence we can see that
the bifurcation surface intersects the asymptotic boundary in the
Poincar\'e coordinates with
 \be
r^2=z^2\sinh^2 u=L^2\tanh^2 u\too{u\to\infty} L^2\,. \labell{connect1}
 \ee
That is, on a sphere of radius $r=L$ in the Minkowski coordinates of
the boundary CFT. This will be the entangling sphere $\Sigma$ in the
calculation of the entanglement entropy. One can work harder to find
the precise connection between the two coordinate systems on the
boundary of AdS$_{d+1}$. The resulting coordinate transformation is
precisely that given in eq.~\reef{trn1} with $R=L$ and
$\ttau=\tau$.\footnote{In fact, examining the asymptotic relation
between the Poincar\'e and hyperbolic foliations is how we originally
derived the transformation \reef{trn1}.}

\FIGURE[b]{  
\includegraphics[width=0.45\textwidth]{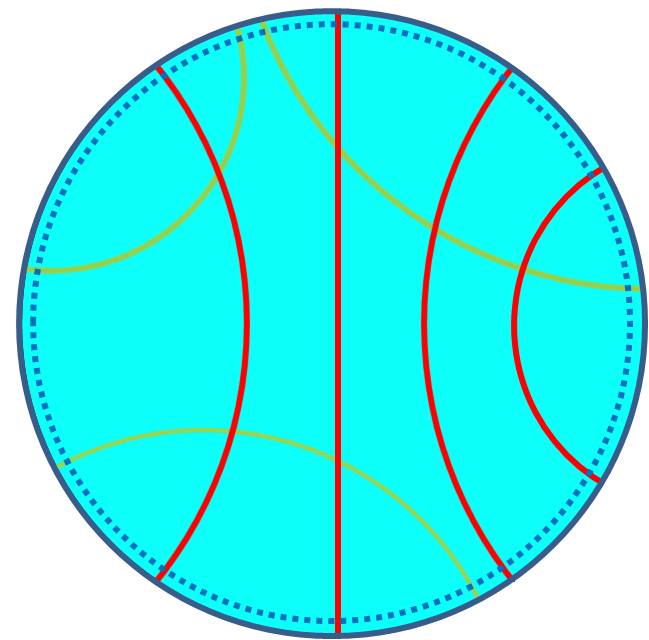}
\caption{A slice of constant $t$ through the AdS$_{d+1}$ spacetime. The
solid (red and green) arcs each represent codimension two surfaces with
constant curvature $-1/L^2$. These are all connected by isometries of
the AdS$_{d+1}$ geometry. The red arcs are connected by a boost, as
described in the main text. The dashed blue line represents a regulator
surface at large radius.} \label{horizons9}}
Next we must determine the Hawking temperature of our topological black
hole \reef{hyperx3}. A straightforward calculation shows that
$T=1/(2\pi L)$. Given that the entangling sphere has a radius $R=L$,
this result precisely matches the temperature found in section
\ref{cft}, where the discussion applied for any CFT without reference
to holography. Hence as discussed in the introductory remarks of this
section, changing coordinates in the bulk between the Poincar\'e and
hyperbolic foliations implements the conformal mapping from $\D$, the
causal development of the sphere at $r=L$, to $R\times H^{d-1}$. As
expected from the general considerations of section \ref{cft}, this
mapping generates a thermal state in the $R\times H^{d-1}$ geometry.
The entanglement entropy across the sphere $r=L$ in flat space equals
the thermal entropy of this Gibbs state and further, in the holographic
setting, the latter is given by the horizon entropy of the bulk black
hole.

We leave the calculation of the entropy for a bit later. First we would
like to extend the previous holographic calculations to a sphere in
Minkowski space with an arbitrary radius $R$, as in eq.~\reef{trn1}.
Essentially we want to move the bifurcation surface above to a new
position that intersects the boundary in a different way, as
illustrated in figure \ref{horizons9}. An interesting observation is
that in fact all of the surfaces illustrated in this figure are
`identical' in that they are connected by an isometry of the
AdS$_{d+1}$ geometry. Hence one finds that the bulk metric for the
hyperbolic foliation in which any of these surfaces appears as the
bifurcation surface is identical to that in eq.~\reef{hyperx3}. Given
these isometries, one might ask how the physics (\eg the entropy) can
change. However, in the AdS/CFT framework, calculations are always made
with reference to a regulator surface in the asymptotic region, \eg
$z=z_{min}$ in the Poincar\'e metric \reef{poincare3}. Below, we will
see that the regulator surface plays an essential role in the
calculation of the entanglement entropy. This surface is typically not
invariant under the isometries that connect the various candidate
bifurcation surfaces and consequently, \eg the entropy will be
different with different choices.

One simple isometry relating different bifurcation surfaces,
illustrated in figure \ref{horizons9}, is a boost in the embedding
space of the $y_A$ coordinates. In particular, a boost in the
$(y_{-1},y_d)$-plane leaves the hyperboloid \reef{hyperbola} invariant
while transforming the coordinates:
 \beqa
y'_{-1}&=&\cosh\beta\,y_{-1}-\sinh\beta\,y_d\,,
 \labell{boost}\\
y'_{d}&=&\cosh\beta\,y_{d}-\sinh\beta\,y_{-1}\,.
 \nonumber
 \eeqa
Now we make the trivial substitution to replace eq.~\reef{hyperx0} with
 \beqa
y'_{-1}&=&\rho\cosh u\,,\quad y_0=\tr\sinh(\ttau/L)\,,\quad
y'_d=\tr\cosh(\ttau/L)\,,\labell{hyperx4}\\
y_1&=&\rho\sinh u\cos\phi_1\,,\ y_2=\rho\sinh u\sin\phi_1\cos\phi_2\,,
\cdots\ y_{d-1}=\rho\sinh u\sin\phi_1\sin\phi_2\cdots\sin\phi_{d-2}\,.
 \nonumber
 \eeqa
As before, the constraint \reef{hyperbola} yields $\tr^2=\rho^2-L^2$
and the induced metric is identical to that given in
eq.~\reef{hyperx3}. We leave our choice of Poincar\'e coordinates
unchanged as in eq.~\reef{poincare}. Hence to relate the two coordinate
patches, it is useful to write
 \beqa
y_{-1}&=&\cosh\beta\,\rho\cosh u
+\sinh\beta\,\sqrt{\rho^2-L^2}\cosh(\ttau/L)\,,
 \labell{hyperx5}\\
y_{d}&=&\cosh\beta\,\sqrt{\rho^2-L^2}\cosh(\ttau/L)
+\sinh\beta\,\rho\cosh u\,.
 \nonumber
 \eeqa

Again, this hyperbolic foliation \reef{hyperx3} lends itself to an
interpretation as a topological black hole. However, the horizon has
been displaced relative to the Poincar\'e coordinate patch. To
understand the relation between the two coordinate patches, we examine
how the bifurcation surface at $\rho=L$ (and any finite value of
$\ttau$) approaches the AdS boundary. Combining eqs.~\reef{poincare}
and \reef{hyperx5}, we find
 \be
 y_{-1}+y_d=\frac{L^2}{z}=e^\beta\,L\cosh u\,, \qquad y_1^2+\cdots+y_{d-1}^2=
 \frac{L^2}{z^2}r^2=L^2\sinh^2 u\,.
 \labell{connect4}
 \ee
Recall that $r^2=(x^1)^2+...+(x^{d-1})^2$. Hence we can see that the
bifurcation surface intersects the asymptotic boundary in the
Poincar\'e coordinates with
 \be
r^2=z^2\sinh^2 u=e^{-2\beta}\,L^2\tanh^2 u\too{u\to\infty}
 e^{-2\beta}L^2\,. \labell{connect5}
 \ee
That is, on a sphere of radius $R=e^{-\beta}L$ in the flat boundary
metric. With a bit of work, one can determine the precise relation
between the two coordinate systems on the boundary of AdS$_{d+1}$. In
fact, one finds precisely the coordinate transformation \reef{trn1}
with $R=e^{-\beta}L$ and $\ttau=e^{-\beta}\tau$.

Interpreting the bulk metric \reef{hyperx3} as a topological black
hole, the boundary CFT on $R\times H^{d-1}$ is in a thermal state.
Since the metric is unchanged, the Hawking temperature of the black
hole will be $\tilde{T}=1/(2\pi L)$, precisely as before. We denote
this temperature as $\tilde T$ since it is determined for energies
conjugate to time translations in $\ttau$. To compare to the discussion
in section \ref{cft}, we wish to determine the temperature $T$
conjugate to the time coordinate $\tau$ in the metric \reef{final}. As
noted above, these two coordinates are related by the scaling
$\ttau=e^{-\beta}\tau$. Hence the temperatures are also related by
$\tilde{T}= e^{-\beta} T$ which yields
 \be
T=\frac{1}{2\pi\, e^{-\beta}L}=\frac{1}{2\pi R}\,.
 \labell{temperx1}
 \ee
Hence we have again reproduced precisely the temperature emerging in
our general considerations of CFT's in section \ref{cft}.

Next we turn to the calculation of the horizon entropy, which we
reduced to eq.~\reef{Waldformula2} in the introductory remarks above.
The expression there is simply proportional to the area of the horizon,
which has the geometry $H^{d-1}$. As already noted, the total horizon
entropy diverges but this is precisely the expected result from section
\ref{therm}. Both the boundary and bulk description yield a uniform
constant for the entropy density and so when integrated over the entire
$H^{d-1}$ geometry, it produces a divergent total entropy. As discussed
in section \ref{therm}, equality of this thermal entropy and the
entanglement entropy requires a certain relation \reef{cutoff2} between
the long distance cut-off introduced to regulate the thermal entropy in
$R\times H^{d-1}$ and the short distance cut-off required to regulated
the entanglement entropy across the sphere in Minkowski space. We now
show that the same relation naturally appears in the holographic
framework. In the Poincar\'e coordinates \reef{poincare3}, a short
distance cut-off $\delta$ in the CFT is implemented by introducing a
minimal radial coordinate $z_{min}$ which cuts off the asymptotic
region of the AdS geometry. The standard holographic dictionary for
these two cut-offs is simply $z_{min}=\delta$. On the horizon (in
particular, on the bifurcation surface), eq.~\reef{connect4} gave the
relation $\cosh u=e^{-\beta}L/z$ and so we find the maximum radius to
regulate the calculation of the horizon entropy is naturally given by
 \be
\cosh u_{max}=\frac{e^{-\beta}L}{z_{min}}=\frac R\delta\,.
 \labell{connect3b}
 \ee
This relation can also be re-written as
 \be
\exp(-u_{max})=\frac{R}{\delta}-\sqrt{\frac{R^2}{\delta^2}-1}\simeq
\frac{\delta}{2R}\,.
 \labell{connect3}
 \ee
Hence, to leading order, we find agreement with the CFT expression in
eq.~\reef{cutoff2}.

We are now prepared to evaluate the horizon entropy
\reef{Waldformula2}. To facilitate a comparison with the results of
\cite{cthem,cthem2}, we first make a change in the radial coordinate on
the hyperbolic plane to $x=\sinh u$. Hence the IR regulator
\reef{connect3b} becomes
 \beqa
x_{max}=\sinh u_{max}&=&\sqrt{\frac{R^2}{\delta^2}-1}
 \labell{newcut}\\
&=&\frac{R}{\delta}\left(1+\frac{1}{2}\frac{\delta^2}{R^2}+\cdots\right)\,.
 \nonumber
 \eeqa
Given the bulk metric \reef{hyperx3}, the horizon entropy
\reef{Waldformula2} becomes
 \be
S=\frac{2\,\Gamma(d/2)}{\pi^{d/2-1}}\, \ads\
\Omega_{d-2}\int_0^{x_{max}} \frac{x^{d-2}\,dx}{\sqrt{1+x^2}}\,,
 \labell{stotal1}
 \ee
where $\Omega_{d-2}=2\pi^{(d-1)/2}/\Gamma((d-1)/2)$ is the area of a
unit $(d-2)$-sphere. This result precisely matches eq.~(4.5) in
\cite{cthem2} (up to the replacement $x\to\rho$).

Hence as in \cite{cthem2}, we observe that the leading contribution
arising from eq.~\reef{stotal1} can be written as
 \be
S\simeq \frac{2\pi}{\pi^{d/2}}\frac{\Gamma(d/2)}{d-2}\,\ads\,
\frac{{\mathcal A}_{d-2}}{\delta^{d-2}}+\cdots\,,
 \labell{stotal2}
 \ee
where ${\mathcal A}_{d-2}=\Omega_{d-2}R^{d-2}$ is the `area' of the
entangling surface, \ie an $S^{d-2}$ of radius $R$. Hence this leading
divergence takes precisely the form expected for the `area law'
contribution to the entanglement entropy in a $d$-dimensional CFT
\cite{ryut1,ryut2}. Further we note that the hyperbolic geometry of the
horizon was essential to ensure the leading power was $1/\delta^{d-2}$
here despite the area integral being ($d\!-\!1$)-dimensional in
eq.~\reef{stotal1}. This divergent contribution to the entanglement
entropy is not universal, \eg see \cite{ryut1,ryut2}. However, a
universal contribution can be extracted from the subleading terms. The
form of the universal contribution to the entanglement entropy depends
on whether $d$ is odd or even. For even $d$, the universal term is
logarithmic in the cut-off while for odd $d$, it is simply a constant
term \cite{ryut1,ryut2}. In the present case, expanding the
entanglement entropy \reef{stotal1} in powers of $R/\delta$, we find
the following universal contributions:
\be S_{univ}=\left\lbrace
\begin{matrix}
(-)^{\frac{d}{2}-1}\, {4}
\, \ads\, \log(2R/\delta)&\quad&{\rm for\ even\ }d\,,\\
(-)^{\frac{d-1}{2}}\,\, {2\pi} \, \ads\ \ \ \ \ \ \ \ \ \ \ &\quad&{\rm
for\ odd\ }d\,.
\end{matrix}\right.
\labell{unis} \ee
Further, recall that for even $d$, $\ads=A$, the coefficient of the
$A$-type trace anomaly in the boundary CFT.

\subsection{Entanglement entropy in a cylindrical background}
\label{static}

We now apply the results of our general CFT story discussion in section
\ref{newstor} to a holographic calculation of entanglement entropy in
an $R\times S^{d-1}$ background. In particular, we choose the
entangling surface to be ($d$--2)-dimensional sphere of a fixed angular
radius $\theta_0$. We begin, as before, with the description of
AdS$_{d+1}$ as a hyperboloid \reef{hyperbola} embedded in $R^{2,d}$. In
this case, we relate the embedding coordinates to the standard global
coordinates on the AdS geometry
 \beqa
y_{-1}&=&\trho\cos(t/L)\,,\quad y_0=\trho\sin(t/L)\,,\quad
y_d=\vrho\cos\theta\,,\labell{glob0}\\
y_1&=&\vrho\sin\theta\cos\phi_1\,,
y_2=\vrho\sin\theta\sin\phi_1\cos\phi_2\,,\ \cdots\
y_{d-1}=\vrho\sin\theta \sin\phi_1\sin\phi_2\cdots\sin\phi_{d-2}\,.
 \nonumber
 \eeqa
In this case, the constraint \reef{hyperbola} yields
$\trho^2=\vrho^2+L^2$ and the induced metric becomes
 \be
 ds^2=\frac{d\vrho^2}{\frac{\vrho^2}{L^2}+1}-\left(\frac{\vrho^2}{L^2}+1\right)dt^2
+\vrho^2\left(d\theta^2+\sin^2\theta\,d\Omega^2_{d-2}\right)\,.
 \labell{glob3}
 \ee
If as usual, we consider the asymptotic limit $\vrho\to\infty$ and
remove a factor of $\vrho^2/L^2$, the boundary metric reduces to
 \be
 ds^2_\mt{CFT}=-dt^2
+L^2\left(d\theta^2+\sin^2\theta\,d\Omega^2_{d-2}\right)\,.
 \labell{glob3b}
 \ee
That is, we are studying the boundary CFT in the background geometry
$R\times S^{d-1}$, where the radius of curvature of the sphere is $L$.

Now we wish to connect this coordinate choice to that giving the
hyperbolic foliation in eqs.~\reef{hyperx4} and \reef{hyperx5}. For
simplicity of the presentation, we again focus on the bifurcation
surface in the corresponding metric \reef{hyperx3}. We approach the
latter by taking $\rho\to L$ with $\ttau$ fixed, which yields $y_0=0$
and in turn we find $t=0$ in eq.~\reef{glob0}. Then using
eq.~\reef{hyperx5}, we find that
 \be
\frac{y_d}{y_{-1}}=\frac{\vrho}{\sqrt{\vrho^2+L^2}}\cos\theta
=\tanh\beta\,. \labell{glob5}
 \ee
Hence in the asymptotic limit $\vrho\to\infty$, we see that the
bifurcation surface reaches an angular size on the $S^{d-1}$ with
 \be
\cos\theta_0=\tanh\beta\quad{\rm or}\quad
\sin\theta_0=\frac{1}{\cosh\beta}\,. \labell{glob6}
 \ee
This sphere is then the entangling surface in our calculation of
entanglement entropy in the boundary CFT. We can further construct the
full relation between the two coordinate systems and the result is
identical to that given in eq.~\reef{trn3} with the substitutions: $R=
L$ and $\tau=\ttau$.\footnote{Again, we originally derived the
transformation \reef{trn3} with this approach.}

The entropy is again given by interpreting the hyperbolic foliation
\reef{hyperx3} as a topological black hole and evaluating the horizon
entropy, as in eq.~\reef{Waldformula2}. However, we must first
determine the appropriate IR regulator to introduce in the integral
over the horizon. Given the global coordinates \reef{glob3}, the
standard AdS/CFT dictionary introduces a short distance cut-off in the
boundary CFT as a maximum radius: $\vrho_{max}=L^2/\delta$. Using the
two expressions for $y_{-1}$ in the two coordinate systems, we find
that the bifurcation surface intersects this regulator surface at
 \be
\cosh u_{max}=\frac{L\sin\theta_0}{\delta}\left(1+
\frac{\delta^2}{L^2}\right)^{1/2}\,.
 \labell{inter0}
 \ee
We can relate this result to the CFT discussion in section
\ref{newstor} as follows. There a small angle $\delta\theta$ was chosen
to regulate the calculation of the entanglement entropy. This small
angle defines a small proper distance on the $S^{d-1}$ and so we relate
this angular regulator to the short distance regulator above with
$\delta=L\delta\theta$. Then a bit of algebra allows us to re-write
eq.~\reef{inter0} as
  \beq
 \exp(-u_{max})
 \simeq\frac{\delta\theta}{2\sin\theta_0}+\cdots\,.
 \labell{cutoff5}
 \eeq
Hence the holographic framework reproduces (to leading order) the
relation \reef{cutoff4} between the UV and IR regulators determined
purely in our general discussion of CFT's.

Our goal is to evaluate the holographic entanglement entropy across the
entangling sphere at $\theta=\theta_0$ which is again found by
calculating the horizon entropy of the topological black hole
\reef{hyperx3}. Hence we return to eq.~\reef{Waldformula2} which we
write as
 \be
S=\frac{2\,\Gamma(d/2)}{\pi^{d/2-1}}\, \ads\
\Omega_{d-2}\int_0^{x_{max}} \frac{x^{d-2}\,dx}{\sqrt{1+x^2}}\,,
 \labell{stotal3}
 \ee
where, as above, we have chosen $x=\sinh u$ as an alternate radial
coordinate on the hyperbolic plane $H^{d-1}$. With eq.~\reef{inter0},
the regulator radius then becomes
 \beq
x_{max}=\sinh u_{max}=\frac{L\sin\theta_0}{\delta} \left(
1-\frac{\delta^2}{L^2}\cot^2\theta_0\right)^{1/2}
 \labell{newcut2}\,.
 \eeq
With this result, we observe that the leading contribution arising from
eq.~\reef{stotal2} reproduces the expected `area law' contribution to
the entanglement entropy \cite{ryut1,ryut2}. The universal
contributions, appearing in the subleading terms \cite{ryut1,ryut2},
now take the form:
\be S_{univ}=\left\lbrace
\begin{matrix}
(-)^{\frac{d}{2}-1}\, {4} \, \ads\,
\log\left(\frac{2L}{\delta}\sin\theta_0\right)&\quad& {\rm for\ even
\ }d\,,\\
(-)^{\frac{d-1}{2}}\,\, {2\pi} \, \ads\ \ \ \ \ \ \ \ \ \ \ &\quad&{\rm
for\ odd\ }d\,.
\end{matrix}\right.
\labell{unis2} \ee
Recall that here $L$ denotes the radius of curvature of the $S^{d-1}$.

This derivation of the holographic entanglement entropy extends the
calculations presented in \cite{cthem,cthem2} which only examined the
special case $\theta_0=\pi/2$, \ie the entangling sphere was chosen to
be the equator of the background $S^{d-1}$ in the cylindrical
background. Of course, setting $\theta_0=\pi/2$ above, we recover the
results derived in \cite{cthem,cthem2}. Note that this agreement in
trivial for odd $d$, since the universal term is independent of
$\theta_0$ (as well as $L$). Further we note that the above expressions
are essentially the same (up to $L\to R$) as those in eq.~\reef{unis}
where the background geometry was just flat Minkowski space.

\section{A CFT calculation} \label{CFT2}

In the previous section, we have found for a broad range of geometries
that the universal contribution to the entanglement entropy for general
holographic CFT's is controlled by the central charge $\ads$. For even
dimensional CFT's, this charge coincides precisely with the coefficient
of the $A$-type trace anomaly. Much of the motivation for this work
came from \cite{cthem,cthem2}, where a similar result was found for the
entanglement entropy between the two halves of the sphere in the
background geometry $R \times S^{d-1}$. In fact, in \cite{cthem2}, it
was shown that their result actually applied for any CFT in even $d$,
without reference to holography. Here we would like to show that our
present results can also be extended to general CFT's.

\subsection{Mapping to de Sitter space}

In order to calculate the entanglement entropy of the sphere for a CFT,
we find it convenient to use a mapping to (the static patch of) de
Sitter space, instead of the mapping to $R \times H^{d-1}$. We start
again with the flat space metric for $d$-dimensional Minkowski space in
polar coordinates:
 \beq
ds^2=-dt^2+dr^2+r^2\,d\Omega^2_{d-2}\,.
 \labell{flat1}
 \eeq
Now with the coordinate transformation
 \beqa
t&=&R\,\frac{\cos\theta\,\sinh(\tau/R)}{1+\cos\theta\,\cosh(\tau/R)}\,,
 \labell{trn2}\\
r&=&R\,\frac{\sin\theta}{1+\cos\theta\,\cosh(\tau/R)}\,,
 \nonumber
 \eeqa
we can readily verify the flat space metric \reef{flat1} becomes
 \bea
ds^2&=&\Omega^2\, \left[-\cos^2\!\theta\, d\tau^2+
R^2\left(d\theta^2+\sin^2\!\theta\,d\Omega^2_{d-2}\right)\right]
 \labell{round0}\\
{\rm where}&&\qquad\Omega=(1+\cos\theta\,\cosh(\tau/R))^{-1}\,.
 \nonumber
 \eea
After eliminating the conformal factor $\Omega^2$, the remaining metric
corresponds to the static patch of $d$-dimensional de Sitter space with
curvature scale $R$. The latter identification may be clearer if we
transform to $\hr= R\sin\theta$, which puts the above metric in the
form
 \beq
ds^2=-\left(
1-\frac{\hr^2}{R^2}\right)\,d\tau^2+\frac{d\hr^2}{1-\frac{\hr^2}{R^2}}
+\hr^2\,d\Omega^2_{d-2}\,.
 \labell{twor}
 \eeq
With eq.~\reef{trn2}, we observe that
 \beqa
\tau\rightarrow\pm\infty\,:&&(t,r)\rightarrow(\pm R,0)\labell{junker9}\\
\theta\rightarrow\frac{\pi}{2}\,:&&(t,r)\rightarrow(0,R)\nonumber
 \eeqa
Note that $\theta=\pi/2$ corresponds to the cosmological horizon at the
boundary of the static patch. In any event, with eq.~\reef{junker9}, we
see that the new coordinates cover precisely the causal development
$\D$ of the ball $r\le R$ on the surface $t=0$.

Recall that inside $\D$, the modular transformations act geometrically
along the flow in eq.~(\ref{fl}). This transformation corresponds
through the conformal mapping (\ref{trn2}) to the time translation
$\tau\rightarrow \tau+ 2 \pi R\,s$ in de Sitter space --- just as
happened with the mapping to $R\times H^{d-1}$. Therefore the modular
transformations act geometrically as time translations in the static
patch and the state in the de Sitter geometry is thermal at temperature
$T=1/(2\pi R)$ with respect to the Hamiltonian $H_\tau$ generating
$\tau$ translations, \ie the density matrix is given by $\rho\sim
\exp\left[-2 \pi R\, H_{\tau}\right]$. Again, this result generalizes
observations made in ref.~\cite{candow}, which examined conformal
mappings of free conformal field theories in $d=4$.

\subsection{Thermodynamic entropy}

As in section \ref{cft}, the entanglement entropy for the sphere of
radius R in flat space is equivalent to the thermodynamic entropy of
the thermal state in de Sitter space. We are going to use standard
thermodynamics in order to compute this entropy. Normalizing the
thermal density matrix $\rho=e^{-\beta H_{\tau}}/\textrm{tr}(e^{-\beta
H_{\tau}})$, we calculate the von Neumann entropy
 \bea
S=-\textrm{tr}(\rho\, \log \rho)&=&\beta \, \textrm{tr} (\rho H_\tau )+
\log \textrm{tr} (e^{-\beta H_\tau})
 \nonumber\\
&=&\beta E\,-\,W \labell{freeenergy}
 \eea
where $W=-\log Z$ denotes the `free energy' of the partition function
$Z=\textrm{tr} \left(\exp[-2 \pi R\,H_\tau]\right)$.

The energy term in (\ref{freeenergy}) is, of course, the expectation
value of the operator which generates $\tau$ translations. Since the
latter translations correspond to a Killing symmetry of the static
patch (\ref{round0}), $E$ is just the Killing energy which can be
expressed as
 \beq
E=\int_V d^{d-1}x \sqrt{h}\, \langle\, T_{\mu \nu}\,\rangle\, \xi^\mu\,
n^\nu= - \int_Vd^{d-1}x  \sqrt{-g}\,  \langle\,
T^\tau{}_\tau\,\rangle\,,
 \labell{energy}
 \eeq
where the integral runs over $V$, a constant $\tau$ slice out to
$\theta=\pi/2$. Further $n^\mu \partial_\mu \equiv
\sqrt{|g^{\tau\tau}|}\,\partial_\tau$ is the unit vector normal to $V$
and $\xi^\mu\partial_\mu\equiv
\partial_\tau$ is the time translation Killing vector.

Recall for the present investigation, the quantum field theory can be
any general CFT. Then the state, coming from a conformal transformation
of the Minkowski vacuum, is invariant under de Sitter symmetry group
\cite{birrell,laflame} and we have
 \beq
\langle\, T^\mu{}_\nu\,\rangle=\kappa \,\delta^\mu{}_\nu\,,
 \labell{stud9x}
 \eeq
where $\kappa$ is some constant. Hence the expectation value of the
stress tensor is completely determined by the conformal anomaly
\reef{traza}. Note that in eq.~\reef{traza}, all of the the Weyl
invariants vanish in de Sitter space, \ie $I_n=0$, while the Euler
density $E_d$ yields a constant depending on the de Sitter radius $R$.
Hence we can fix the constant $\kappa$ in eq.~\reef{stud9x} as
 \beq
\langle\, T^\mu{}_\nu\,\rangle=-2\,(-)^{d/2}A\, \frac{E_d}d
\,\delta^\mu{}_\nu\,,
 \labell{trace5}
 \eeq
for $d$ even. The energy (\ref{energy}) then becomes
\begin{equation}
E=2\,(-)^{d/2}A\, \frac{E_d}d\, R^{d-1}\Omega_{d-2}\, \int_0^{\pi/2}
d\theta\,\cos\theta\,\sin^{d-2}\theta\,.
 \labell{trace6}
\end{equation}
Hence the energy is finite. However, we are only interested in the
universal coefficient of the logarithmic term in the entropy and hence,
given this result, we may discard the energy contribution in
eq.~\reef{freeenergy}. Note that for $d$ odd, the trace anomaly
vanishes and so we would find $E=0$.

It remains to compute the contribution $W=-\log \textrm{tr}
\left(\exp[-2 \pi R\,H_\tau]\right)$ in eq.~\reef{freeenergy}. This can
be done as usual passing to imaginary time $\tau_\mt{E}$ and
compactifying the Euclidean time with a period $\beta=2\pi R$. The
metric becomes
 \beq
ds^2=\cos^2\!\theta\, d\tau^2+
R^2\left(d\theta^2+\sin^2\!\theta\,d\Omega^2_{d-2}\right) \,.
 \labell{none1}
 \eeq
This Euclidean manifold is precisely a $d$-dimensional sphere with
radius of curvature $R$.\footnote{The fact that this metric
\reef{none1} corresponds to the sphere may be more evident after the
coordinate transformation \cite{fursaev}: $\sin \theta=\sin \theta_1
\sin \theta_2$ and $\tan (\tau/R) = \cos \theta_2 \tan \theta_1$, which
transforms the metric to
 \beq
ds^2=R^2\left(d\theta_1^2+\sin^2 \theta_1^2 d\theta_2^2 + \sin^2
\theta_1 \sin^2 \theta_2 d\Omega^2_{d-2}\right)\,.\nonumber
 \eeq} Note that the periodicity $\Delta
\tau=2\pi R$ is precisely that required to avoid a conical singularity
at $\theta=\pi/2$. Thus, one ends up with the Euclidean path integral
on $S^d$.

Recall that we only need to determine the coefficient of the
logarithmic term in the entropy for even $d$. Now the free energy has a
general expansion
\begin{equation}
W=-\log Z=(\textrm{non-universal terms}) +a_{d+1}
\log{\delta}+(\textrm{finite terms})\,,
 \labell{polo}
\end{equation}
where $\delta$ is our short distance cut-off and the non-universal
terms diverge as inverse powers of $\delta$.  The coefficient $a_{d+1}$
for a conformal field theory is determined by the integrated conformal
anomaly \cite{vassi} --- for free fields, it is one of the coefficients
in the heat kernel expansion. In order to see this, consider an
infinitesimal rescaling of the metric $g^{\mu\nu}\rightarrow(1-2 \delta
\lambda)g^{\mu\nu}$. Since\footnote{Recall we have made the transition
to a Euclidean signature here.}
\begin{equation}
\frac{2}{\sqrt{g}}\frac{\delta W}{\delta g^{\mu\nu}}=\langle\,
T_{\mu\nu}\,\rangle+(\textrm{divergent terms})\,,
 \labell{none2x}
\end{equation}
in terms of the renormalized stress tensor $\langle\, T_{\mu\nu}
\,\rangle $, we have
\begin{equation}
\frac{\delta W}{\delta \lambda}=-\int d^dx\, \sqrt{g}\,
\langle\, T^\mu{}_\mu\,\rangle+(\textrm{divergent terms})\,,
\labell{none3}
\end{equation}
which is the integrated trace anomaly. On the other hand, due to the
conformal invariance of the action, scaling the metric as above must
give the same result as keeping the metric constant but shifting the UV
regulator:  $\delta\rightarrow (1-\delta \lambda)\delta$. Combining
these expressions with eq.~(\ref{polo}), one finds
\begin{equation}
a_{d+1}= \int d^dx\, \sqrt{g}\,\langle\,T^\mu{}_\mu\,\rangle\,.
 \labell{titres}
\end{equation}

Hence we are left to substitute eq.~(\ref{traza}) for the trace anomaly
and integrate over the $S^d$. Here we also need to observe that since
the sphere is conformally flat, all of the Weyl invariants $I_n$ vanish
for the sphere, while our convention in eq.~\reef{traza} is that the
integral of the Euler density on $S^d$ yields $2$. Hence for any CFT in
even dimensions, the universal contribution to the entanglement entropy
becomes
 \beq
S_{univ}= (-1)^{\frac{d}2-1} 4\, A\, \log (R/\delta)\,,
 \labell{none4}
 \eeq
which is essentially the same as our holographic result in
eq.~(\ref{unis}). In particular, we see that the coefficient of the
universal term in the entanglement entropy is proportional to the
central charge $A$. Note that this result \reef{none4} and
eq.~\reef{unis} do not quite agree on the argument of the logarithm. In
eq.~\reef{none4}, $R$ was simply inserted as the only available scale
in the problem whereas the result in eq.~\reef{unis} emerged from a
detailed evaluation of the entropy. Hence the mismatch is no surprise.
However, the difference between the two expressions can be simply
regarded as a non-universal constant term.

Although we do not present the details here, it is straightforward to
extend our analysis above to the entanglement entropy across a sphere
for CFT's in a cylindrical background $R \times S^{d-1}$. The result
for the universal contribution for even $d$ again matches our
holographic result \reef{unis2}. In particular, the coefficient is
controlled by the coefficient of the $A$-type trace anomaly and in
fact, it is identical to that just above in eq.~\reef{none4}. As above,
the present approach would not naturally reveal precisely the same
scale over $\delta$ in the argument of the logarithm, as we found in
eq.~\reef{unis2}.

In the odd dimensional case, we have seen $E=0$ and so
eq.~\reef{freeenergy} reduces to $S=\log Z$. That is, the entanglement
entropy is simply minus the free energy on a sphere. This case for odd
dimensions has recently been examined in \cite{dowker2} for the
particular case of a free scalar field and the results there are in
agreement with this identification.

\subsection{Hyperbolic mapping} \label{hypmap}

It is interesting to go through a similar CFT analysis for our mapping
\reef{trn1} from flat space to $R\times H^{d-1}$. As discussed in
section \ref{therm}, in the latter hyperbolic space, we have a thermal
state with $T=1/(2 \pi R)$. Now, however, this space is not maximally
symmetric and so by symmetry, the stress tensor is only restricted to
have a form
 \begin{equation}
 T^\mu{}_\nu=\textrm{diag}(-\E,\,p,\,\cdots,\,p)\,,
 \labell{none5}
 \end{equation}
with $\E$ and $p$ constants. The background geometry is again
conformally flat and so the Weyl invariants $I_n$ vanish. Further, the
background is the direct product of two lower dimensional geometries
which dictates that the Euler density is also zero. Hence the trace
anomaly \reef{traza} vanishes in this particular background and in
eq.~\reef{none5}, we must have
\begin{equation}
\E=(d-1)\, p\,. \labell{none6}
\end{equation}
To proceed further we must focus on $d=4$ since in general, we do not
know the value of the energy density. However, for $d=4$, we have the
Bunch-Davies-Brown-Cassidy formula \cite{bdc} which relates the stress
tensor for a given state in flat space to that in another space
obtained by a conformal mapping from flat space. With this expression,
we obtain
 \beq
\E=\frac{3a+c}{8 \pi^2 R^4}\,. \labell{fourdx}
 \eeq
Here, we have adopted the standard notation for the central charges in
four-dimensional CFT's, \ie comparing to eq.~\reef{traza}, $a=A$ and
$c=16\pi^2 B_1$ \cite{cthem}.\footnote{Our conventions are such that
these coefficients are normalized to $a=1/360$ and $c=1/120$ for a free
conformally coupled massless (real) scalar.}

Now, in contrast to the discussion in the previous section, the energy
contribution in eq.~\reef{freeenergy} is divergent due to the infinite
volume of the $H^{d-1}$. Using eq.~\reef{fourdx} for $d=4$, we find  a
contribution to the logarithmic term in the entropy
 \bea
2\pi R E&=&8\pi^2 R^4\,\E\,\int_0^{u_{max}} du\,\sinh^2(u)
 \labell{terra}\\
&=& (\textrm{non-universal terms}) +\frac{1}{2}\left(3a+c\right)\,\log
\delta+(\textrm{finite terms})\,.
 \nonumber
  \eea
The final result above relies on choosing the maximum radius $u_{max}$
as in eq.~(\ref{cutoff2}).

The second contribution to the logarithmic term in the entropy
\reef{freeenergy} comes from the free energy $W$ which should again be
determined by the conformal anomaly, as in eqs.~\reef{polo} and
\reef{titres}. However, as noted above, $\langle\,
T^\mu{}_\mu\,\rangle=0$ in the present case and so we have a zero
`bulk' contribution from the anomaly. However, implicitly, the manifold
comes with a boundary, \ie $u=u_{max}$ in the cut-off manifold. Hence
the conformal anomaly should pick up boundary contributions there but
unfortunately, at present, the expression for these terms is unknown.
One could take the results for the entropy \reef{none4} and the energy
\reef{terra} in $d=4$ to determine this boundary contribution as
 \be
W = (\textrm{non-universal terms}) +
\frac{1}{2}\left(-5a+c\right)\,\log \delta \,.
 \labell{none7}
 \ee
In particular, the term proportional to $c$ in eq.~(\ref{terra}) must
cancel with a boundary term. It is intriguing that this approach may
imply that there are further restrictions for the boundary conditions
that should be used to define the cut-off at $u=u_{max}$.

\section{Discussion} \label{discuss}

To summarize our results, we have produced a derivation of holographic
entanglement entropy for certain geometries, namely, with spherical
entangling surfaces. The derivation started by finding an appropriate
calculation of the entanglement entropy in the boundary CFT in section
\ref{cft}. Here, we have avoided the usual approach of using the
replica trick \cite{cardy0,callan}. Rather we used conformal
transformations to relate the entanglement entropy across a spherical
entangling surface to the thermal entropy in a new background geometry,
$R\times H^{d-1}$. While this construction applies for any CFT, it is
not particularly useful in general as it simply relates two difficult
problems to one another. However, in the case of a holographic CFT, the
AdS/CFT correspondence translates the second problem to the question of
determining the horizon entropy of a topological black hole, as
described in section \ref{ads}. The latter is a straightforward
calculation using Wald's entropy formula \reef{Waldformula}. Hence we
have derived an expression for the holographic entanglement entropy
\reef{Waldformula2}, which applies for any bulk gravitational theory,
albeit for the specialized case of a spherical entangling surface. We
have explicitly considered the entanglement entropy for
flat space, in section \ref{flat2}, or for a cylindrical background, in
section \ref{static}. This discussion would also straightforwardly
extend to a spherical entangling surface in a background $R\times
H^{d-1}$. The key difference, however, would be that we would not start
with the vacuum state in this background, rather we take the thermal
state that is equivalent to the vacuum in $R^d$ by the conformal
mappings introduced in section \ref{cft}.

The present discussion extends the derivation presented in
\cite{cthem2}, where the analysis focused on a special case of the
geometries considered here. There the background geometry was chosen to
be $R\times S^{d-1}$ and the entangling surface was placed on the
equator of the $S^{d-1}$. We might comment, however, that the enhanced
symmetry of the latter geometry allowed for an alternate derivation
which was based on the replica trick.

In section \ref{cft}, both the entanglement and thermal entropy were
divergent and their equality was only guaranteed by imposing a relation
between the short-distance cut-off required to regulate the
entanglement entropy and the long-distance cut-off used to regulate the
thermal entropy, as in eqs.~\reef{cutoff2} and \reef{cutoff4}. Hence
the conformal mapping introduced an interesting UV/IR relation between
the two states of the CFT, which is reminiscent of the UV/IR connection
found in the AdS/CFT correspondence. The holographic dictionary also
naturally reproduced these relations in section \ref{ads}. In this
case, the horizon of the topological black hole extended out to the AdS
boundary and the desired relations were determined examining the
intersection of the horizon with the UV regulator surface associated
with the original background geometry. The intersection of these two
surfaces was also the key feature which distinguished different
horizons connected to boundary spheres with different sizes. Otherwise
the horizons are `identical', in that, these surfaces can all be mapped
into one another by an isometry of the AdS$_{d+1}$ geometry.

Examining the results in eqs.~\reef{unis} and \reef{unis2}, we observe
that the universal contribution to the entanglement entropy is
proportional to the central charge, $\ads$, which characterizes the
boundary CFT. This charge was introduced in \cite{cthem,cthem2} where
it was observed that this charge satisfies a holographic
c-theorem\footnote{See also recent related results in
\cite{friends2,jimt,aninda6,miguel99x}.} --- a result which was
conjectured to extend to general field theories. Following the
considerations of the holographic principle in \cite{lenny}, it was
also argued that $\ads$ gave a measure of the number of degrees of
freedom in the boundary field theory \cite{cthem2}. Given the present
results, we can identify this charge using entanglement entropy for a
broader class of geometries, in particular, if we wish to examine the
c-theorem noted above outside of a holographic framework. Specifically,
with eq.~\reef{unis}, $\ads$ can always be identified using a spherical
entangling surface in flat space.

For the case of an even dimensional boundary theory (\ie $d$ even), the
central charge is precisely that appearing in the $A$-type trace
anomaly \reef{traza}. In this case, the universal term in the
entanglement entropy is proportional to a logarithm of the cut-off
scale $\delta$. Of course, the latter is balanced by another scale to
make the argument of the logarithm dimensionless. In the flat Minkowski
background, this scale is naturally set by the size of the entangling
sphere, as in eq.~\reef{unis}, since this is the only scale in the
construction. In our result \reef{unis2} for the Einstein universe
background, the other scale is $L\sin\theta_0$ (up to a factor of 2,
which also appears in the flat space calculation). This scale can be
readily identified as the radius of curvature of the entangling sphere.
Hence, in this sense, the same scale appears in the result for both of
our calculations.

We also note that our result \reef{unis2} is a simple generalization of
the standard result for the entanglement entropy of a two-dimensional
CFT. Given a sub-system of length $\ell$ in a full system of size $C$
(with periodic boundary conditions), the entanglement entropy is given
by\cite{cardy0,finn}
 \be
S=\frac{c}{3}\,\log\left(\frac{C}{\pi\delta}\,\sin\frac{\pi\ell}{C}
\right)\,,
 \labell{standx}
 \ee
where $c$ is the central charge of the $d=2$ CFT.\footnote{In general,
one might also expect a non-universal constant term to appear on the
right-hand side.} Of course, for two dimensions, the cylindrical
background considered in section \ref{static} reduces to $R\times S^1$.
Further our result \reef{unis2} precisely matches the result given
above using the relations: $\ads=A=c/12$, $L=C/2\pi$ and
$\theta_0=\pi\ell/C$, which are applicable for $d=2$. Our result in
eq.~\reef{unis2} provides a natural extension of the above expression
for $d=2$ to higher (even) dimensions.

Of course, in section \ref{CFT2}, we were able to show that the
universal contribution to the entanglement entropy takes the same form
as our holographic result \reef{unis} for any CFT in any even number of
dimensions. A similar result was proven in \cite{cthem2} where the
entangling surface was the equator of the $S^{d-1}$ in the static
Einstein universe background. While we did not present the details, it
would be straightforward to extend the arguments of section \ref{CFT2}
to cover this case or a spherical entangling surface of any angular
size. Hence, for even $d$ (without any reference to holography), we
have established that the universal contribution to the entanglement
entropy across a spherical entangling sphere is proportional to $A$.

The calculation in section \ref{CFT2} begins with a mapping of the
causal development $\D$ to the static patch in de Sitter space. This
approach is similar to that in \cite{stud3}. However, in the latter,
 the thermodynamical entropy is written as
 \beq
S=\int^{T=\frac{1}{2 \pi R}}_0
\left.\frac{dE(T,V)}{T}\right\vert_{V=\textrm{const.}}\,.
 \labell{grg}
 \eeq
The main difference\footnote{This approach must also assume that the
entropy at zero temperature vanishes or at least the relevant
logarithmic contribution vanishes. This assumption may be related to
the mismatch in \cite{stud3}, where the coefficient to the logarithmic
term is not given by the $A$-type anomaly for a vector field.} with our
approach is that in eq.~(\ref{grg}), one needs to know the energy `off
shell', \ie away from the point $T=1/(2 \pi R)$. This thermal energy is
known only for the case of free fields and so the results in
\cite{stud3} are only determined for free fields. In contrast, our `on
shell' approach only makes reference to the energy and free energy at
$T=1/(2\pi R)$ which naturally emerges from the conformal
transformation of the Minkowski vacuum. This essential difference
allowed us to derive a general result for any CFT in any even number of
dimensions. However, we might add that it seems a knowledge of the
whole function $E(T,V)$ would certainly be necessary for computing the
Renyi entropies $S_n=(1-n)^{-1}\log(\textrm{tr}\rho^n)$ for general
$n$.

The general connection of the universal terms in entanglement entropy
and the trace anomaly was first noted in \cite{finn} for $d=2$ and
extended to higher dimensions in \cite{ryut1,solo1}. In particular,
\cite{solo1} established that the entanglement entropy for a spherical
entangling surface would be proportional to $A$ in four dimensions.
However, we must comment that the arguments presented in
\cite{ryut1,solo1} are only completely justified when there is a
rotational symmetry in the transverse space around the entangling
surface. For configurations without this symmetry, additional
correction terms must be added to the entanglement entropy, however,
they still seem to have the same general character, \ie they are
geometric expressions evaluated on the entangling surface with
coefficients linear in the central charges \cite{mishanet,solo1,adam}.

We note that the coefficient of the universal contribution is identical
for any sphere of any size in flat space or in a cylindrical
background. Given that all of these geometries are related by conformal
mappings, this reflects the fact that this coefficient is conformally
invariant, as noted in \cite{dowker1}, and clarified in the present
paper. Another case then, which also refers to a spherical entangling
surface was recently discussed in \cite{solo2}. This case is the
near-horizon geometry of an extreme black hole, which has a geometry
$H^2\times S^{d-2}$ and by explicit calculation for a free conformally
coupled scalar in any even dimension, it was shown that the coefficient
of the log term was controlled by $A$. The event horizon in this
geometry can be conformally mapped to a sphere in flat space. Hence one
expects the coefficient for the logarithmic correction to the black
hole entropy is also controlled by the $A$-type trace anomaly for even
$d$. In fact, given the conformal invariance of this coefficient, the
discussion in \cite{cthem2} would be sufficient to indicate that $A$
also appears as the coefficient of the universal contribution for any
CFT in geometries considered in section\ref{CFT2}.

Our calculations of holographic entanglement entropy also yielded
results for odd $d$ in eqs.~\reef{unis} and \reef{unis2}. In this case,
following \cite{ryut1}, we have identified the universal contribution
as the constant term appearing in the expansion in powers of the
cut-off. Hence the result is completely independent of the size of the
entangling sphere or the background geometry. The universality of this
constant contribution to the entanglement entropy is established for a
variety of three-dimensional systems \cite{wenx,fradkin}. However, we
should note that one may worry that in general the precise value of
this constant will depend on the details of the regulator --- as
discussed in \cite{cthem2}.\footnote{RCM thanks M. Smolkin, A.
Schwimmer and S. Theisen for discussions on this point.} We expect that
these issues can be circumvented by considering an appropriate
construction with mutual information --- \eg see
\cite{casini55,swingle}. When calculated with two separate regions, the
latter is free of divergences and any regulator ambiguities.

Recall our result in section \ref{CFT2} for the case of odd dimensional
CFT's. Namely, the entanglement entropy for a spherical entangling
surface $S^{d-2}$ is precisely minus the free energy of the CFT on a
sphere $S^d$, \ie we have
 \be
 S=\log Z\qquad{\rm for\ odd}\ d\,.
 \labell{jack}
 \ee
This result can be related to a calculation of entanglement entropy
described in \cite{cthem2}. There, the initial problem was to determine
the entanglement entropy of a $d$-dimensional CFT on $R\times S^{d-1}$
when the entangling surface $\Sigma$ was chosen to be the equator of
the sphere. The approach taken was to apply the geometric approach to
the replica trick \cite{callan}, where one evaluates the partition
function on the background geometry with an infinitesimal conical
defect at $\Sigma$. However, this procedure is only well-defined if
there is a rotational symmetry about this surface and so in
\cite{cthem2}, the $R\times S^{d-1}$ background was mapped to $S^d$
using a conformal transformation. The expression for the entanglement
entropy then becomes
  \be
S=\lim_{\epsilon\to0}\left(\frac{\partial\ }{\partial \epsilon}
 +1\right)\,\,\log Z_{1-\epsilon}\,.
  \labell{ees4}
  \ee
where the partition function is evaluated on a `$(1-\epsilon)$-fold
cover' of the $d$-dimensional sphere, with an infinitesimal conical
defect $\Delta\theta=2\pi(1-\epsilon)$ at $\Sigma$. In \cite{cthem2},
this construction was applied to determine the universal contribution
to the entanglement entropy for even $d$. However, it applies just as
well for the case of odd $d$. Hence comparing to eqs.~\reef{jack} and
\reef{ees4}, we see that the leading variation of the sphere partition
function must vanish in the latter equation, \ie $\partial_\epsilon
Z|_{\epsilon=0}=0$. This vanishing of the variation of $Z$ is analogous
to the vanishing of the energy contribution in eq.~\reef{freeenergy}
for odd $d$.

We should note that the expressions on both sides of eq.~\reef{jack}
are expected to diverge and so, as stressed in section \ref{cft}, care
must be taken in applying a consistent regulator to ensure the equality
of the two quantities. As noted before, this equality was established
for free conformal scalar fields in odd dimensions by \cite{dowker2}.
There, in fact, the heat kernel regulator completely eliminated the
divergences on both sides of eq.~\reef{jack}. With a general regulator,
where divergent terms still appear, one still expects that $S$ and
$\log Z$ will be equal order by order in the cut-off scale. In
particular, the universal constant contribution to the entanglement
entropy must match the constant term in free energy.

This last observation is of interest in connection to a recent
discussion of ${\mathcal N}=2$ superconformal field theories in three
dimensions in \cite{jaff}. There the author provides evidence that, for
these theories, the sphere partition function plays a very similar role
to the central charge $a$ in four-dimensional theories with ${\mathcal
N}=1$ supersymmetry. In particular, as function of possible trial
$R$-charges, (the finite part of) $Z$ is extremized by the exact
superconformal $R$-charge, in analogy to $a$-maximization in four
dimensions \cite{ken}. Further, given the connection between
$a$-maximization and the c-theorem for the corresponding field theories
\cite{ken2}, one is naturally lead to speculate that $Z$-maximization
may provide a framework to develop the analog of c-theorem for
supersymmetric theories in three dimensions \cite{jaff}. Now, the
identity in eq.~\reef{jack} connects this suggestion to the broader
conjecture of \cite{cthem,cthem2}. Motivated by holographic evidence,
the authors there proposed that the central charge $\ads$, which
appears in the universal contribution to the entanglement entropy,
should evolve monotonically under RG flows. Focussing on three
dimensions, eq.~\reef{jack} would yield
 \be
\left.\log Z\right|_{finite}=S_{univ}=-2\pi\,a^*_3\,.
 \labell{threeddd}
 \ee
Hence the holographic results of \cite{cthem,cthem2} indicate that any
$d=3$ supersymmetric gauge theories with a gravity dual will satisfy
the desired c-theorem. Alternatively, if the field theoretic approach
of \cite{jaff} can be extended to establish a c-theorem as a
consequence of $Z$-maximization, this would provide further evidence
for the general c-theorem conjectured in \cite{cthem,cthem2} for
quantum field theories in any (odd) number of spacetime dimensions

Given our new results, it is interesting to consider some explicit
applications, in particular, to consider higher curvature corrections
in the holographic entanglement entropy in various string models. A
well-known set of corrections appear at order (curvature)$^4$ in all
superstring theories \cite{curv4}. The supersymmetric completion of
this term in type IIb string theory was used to explicitly construct
all of the interactions involving the curvature and the Ramond-Ramond
five-form \cite{miguel}. These interactions are all naturally written
in terms of the Weyl tensor and certain tensors constructed from the
five-form. Now in a holographic setting, we are considering a
supersymmetric reduction of the form AdS$_5\times \cM_5$ and in such a
background, both the Weyl tensor and the relevant tensors for the
five-form vanish. Hence these interactions do not modify the background
geometry nor do they contribute to the Wald entropy \reef{Waldformula}
of the topological black hole. That is, these particular higher order
corrections to the superstring action leave the entanglement entropy
unchanged. Note that implicitly we have extended the discussion to the
full ten dimensions of the superstring theory and so the horizon has
the geometry $H^3\times \cM_5$. We could first reduce the
ten-dimensional theory to five dimensions, following \cite{univres},
and present the argument entirely in the context of AdS$_5$ as in the
main text. Of course, the result is unchanged.

Further, this result, namely that the entanglement entropy was
unchanged, should have been expected. These $R^4$ interactions appear
at order $\alpha'^3$ with a tree-level and one-loop (\ie $g_s^2$)
contribution. From the perspective of the boundary theory then, these
terms will introduce corrections of order $1/\lambda^{3/2}$ and
$\lambda^{1/2}/N_c^2$ \cite{order}. In particular then, these depend on
the 't Hooft coupling $\lambda$. However, our analysis indicates that
the universal contribution to the entanglement entropy should be
proportion to the central charge $A$ (which is commonly denoted $a$ in
four dimensions). Further in a superconformal gauge theory, it is known
that the central charges are independent of the gauge coupling
\cite{nochange}. Hence, this universal contribution should not receive
any corrections depending on the 't Hooft coupling, which is in accord
with our gravity calculations. We might add that for $N=4$
super-Yang-Mills theory in the limit of zero coupling (\ie the free
field limit), numerical calculations \cite{numerical} of the
entanglement entropy for a spherical entangling surface in flat space
explicitly confirm that the results match the strong coupling result.
Hence these calculations also confirm the same independence of the
gauge coupling.

There are other interesting string theory models where
curvature-squared terms arise in a holographic context \cite{alex}.
These terms originate from the presence of D-branes in the construction
of these backgrounds \cite{ddbrane}. We do not present the details here
but we comment that the presence of these terms does effect the two
central charges, $a$ and $c$, of the dual CFT. In particular, the
difference $c-a$ is controlled by the coefficient of this higher
curvature interaction. However, we are again discussing these terms in
the context where they appear in a controlled perturbative expansion in
string theory and hence they can be modified by field redefinitions. In
particular then, if this interaction is written as $R_{abcd}R^{abcd}$,
it contributes by modifying both the background curvature of the
AdS$_5$ and the expression for the central charge $A$. Of course, this
term also contributes to the Wald entropy and the modifications are
consistent with are final result \reef{unis} and \reef{unis2} where $A$
appears in the coefficient of the universal contribution to the
entanglement entropy. However, with field redefinitions, the
interaction can also be written as $C_{abcd}C^{abcd}$, in which case,
neither the background curvature nor the central charge $a$ are
modified. Further, this term will not contribute to the Wald entropy
and so again the results are consistent with our expressions for the
entanglement entropy.

One issue which our discussion highlights is the close connection
between holographic entanglement entropy and black hole entropy. Indeed
an eternal AdS black hole contains two asymptotically AdS regions and
it has been argued that in this case the horizon entropy corresponds to
the entanglement entropy between the CFT on one boundary and its
thermofield double on the other boundary \cite{juanbh}. Our
construction essentially uses this interpretation for the topological
black hole which corresponds to a pure AdS spacetime. The key
difference from the usual interpretation is that the two asymptotically
AdS regions are complementary portions of the same AdS geometry. It
would be interesting to understand if a similar interpretation is
possible for topological black holes which are endowed with charge
charge and/or rotation \cite{rotate}. Another useful direction would be
to investigate whether AdS space can be foliated in other ways to
produce `topological black holes' with different horizon geometries.
These may then form the basis of a derivation of the holographic
entanglement entropy for entangling surfaces with new geometries.

Of course, our derivation puts the standard proposal of
\cite{ryut0,ryut1} on a firmer footing since we find agreement with
eq.~\reef{define} when the bulk theory is just Einstein gravity. We
might note that the present calculations do not seem to involve the
extremization of some functional over a family of bulk surfaces.
However, given our results, a natural guess might be that when the bulk
gravity theory includes higher curvature terms, we should extend the
definition of holographic entanglement entropy \reef{define} to
extremize the Wald entropy \reef{Waldformula} evaluated on bulk
surfaces homologous to the boundary region of interest. Unfortunately,
one can easily show that this procedure does not produce the correct
entanglement entropy in general, however, interesting progress has
still been made for certain classes of higher curvature theories
\cite{mishanet,friends2}. At the same time, we can add that for generic
entangling surfaces, the bulk surface determining the holographic
entanglement entropy will not play the role of the event horizon for
some black hole \cite{mishanet,friends2}.

Again, our derivation of holographic entanglement entropy discards the
replica trick and instead we are relying on invariance of the
entanglement entropy of the boundary CFT under conformal mappings. Of
course, it would be interesting to extend our approach to produce a
derivation for more general geometries, \ie non-spherical entangling
surfaces in different background spacetimes. A common feature of the
conformal mappings in sections \ref{flat2} and \ref{static}, which
seems important, is that the entangling sphere is mapped to space-like
infinity in $R\times H^{d-1}$. While similar transformations mapping
the entangling surface to infinity are easily constructed for other
geometries, \eg $S^{d-2-n}\times R^n$, the resulting background is
typically time-dependent and it is not evident what the state of the
CFT is in the new background. Hence if further progress is to be made
with this approach, additional insights will be needed with respect to
the most useful conformal mapping to apply for a given geometry. It may
be instructive to translate the discussion of section \ref{CFT2} to a
holographic derivation of the entanglement entropy of a spherical
entangling surface. In any event, there remain many interesting
questions to explore with regards to holographic entanglement entropy.

\vskip 2cm

\noindent {\bf Acknowledgments:} RCM thanks Stuart Dowker, Ben
Freivogel, Janet Hung, Alex Maloney, Aninda Sinha, Misha Smolkin, and
Lenny Susskind for useful discussions. HC is grateful to the Perimeter
Institute for hospitality during the initial stages of this work.
Research at Perimeter Institute is supported by the Government of
Canada through Industry Canada and by the Province of Ontario through
the Ministry of Research \& Innovation. RCM also acknowledges support
from an NSERC Discovery grant and funding from the Canadian Institute
for Advanced Research. MH and HC acknowledge support from CONICET and
Universidad Nacional de Cuyo, Argentina.


\begin{thebibliography}{99}

\bibitem 
 {wenx} See, for example:\\
M.~Levin and X.~G.~Wen,
  ``Detecting Topological Order in a Ground State Wave Function,''
  Phys.\ Rev.\ Lett.\  {\bf 96}, 110405 (2006)
 [arXiv:cond-mat/0510613];\\
A.~Kitaev and J.~Preskill,
  ``Topological entanglement entropy,''
  Phys.\ Rev.\ Lett.\  {\bf 96}, 110404 (2006)
  [arXiv:hep-th/0510092].

\bibitem 
 {cardy0} See, for example:\\
P.~Calabrese and J.~L.~Cardy,
  ``Entanglement entropy and quantum field theory,''
  J.\ Stat.\ Mech.\  {\bf 0406}, P002 (2004)
  [arXiv:hep-th/0405152];\\
P.~Calabrese and J.~L.~Cardy, ``Entanglement entropy and quantum field
theory: A non-technical introduction,''
  Int.\ J.\ Quant.\ Inf.\  {\bf 4}, 429 (2006)
  [arXiv:quant-ph/0505193].

\bibitem 
 {fradkin} B.~Hsu, M.~Mulligan, E.~Fradkin and E.A.~Kim,
``Universal entanglement entropy in 2D conformal quantum critical
points,'' Phys.\ Rev.\ B {\bf 79}, 115421 (2009) [arXiv:0812.0203].

\bibitem 
 {ryut0} S.~Ryu and T.~Takayanagi,
  ``Holographic derivation of entanglement entropy from AdS/CFT,''
  Phys.\ Rev.\ Lett.\  {\bf 96}, 181602 (2006)
  [arXiv:hep-th/0603001].

\bibitem 
 {ryut1} S.~Ryu and T.~Takayanagi,
  ``Aspects of holographic entanglement entropy,''
  JHEP {\bf 0608}, 045 (2006)
  [arXiv:hep-th/0605073].

\bibitem 
 {igor0} I.~R.~Klebanov, D.~Kutasov and A.~Murugan,
  ``Entanglement as a Probe of Confinement,''
  Nucl.\ Phys.\  B {\bf 796}, 274 (2008)
  [arXiv:0709.2140 [hep-th]].

\bibitem 
 {mvr} M.~Van Raamsdonk,
  ``Comments on quantum gravity and entanglement,''
  arXiv:0907.2939 [hep-th];\\
M.~Van Raamsdonk,
  ``Building up spacetime with quantum entanglement,''
  Gen.\ Rel.\ Grav.\  {\bf 42}, 2323 (2010)
  [arXiv:1005.3035 [hep-th]].

\bibitem 
 {head1} M.~Headrick,
  ``Entanglement Renyi entropies in holographic theories,''
  arXiv:1006.0047 [hep-th].

\bibitem 
 {fur0} D.~V.~Fursaev,
  ``Proof of the holographic formula for entanglement entropy,''
  JHEP {\bf 0609}, 018 (2006)
  [arXiv:hep-th/0606184].

\bibitem 
 {ryut2} T.~Nishioka, S.~Ryu and T.~Takayanagi,
  ``Holographic Entanglement Entropy: An Overview,''
  J.\ Phys.\ A  {\bf 42}, 504008 (2009)
  [arXiv:0905.0932 [hep-th]].

\bibitem 
 {mishanet} L.~Y.~Hung, R.~C.~Myers and M.~Smolkin,
  ``On Holographic Entanglement Entropy and Higher Curvature Gravity,''
  arXiv:1101.5813 [hep-th].

\bibitem 
 {callan} C.~G.~Callan and F.~Wilczek,
  ``On geometric entropy,''
  Phys.\ Lett.\  B {\bf 333}, 55 (1994)
  [arXiv:hep-th/9401072].

\bibitem 
 {cthem} R.~C.~Myers and A.~Sinha,
  ``Seeing a c-theorem with holography,''
  Phys.\ Rev.\  D {\bf 82}, 046006 (2010)
  [arXiv:1006.1263 [hep-th]].

\bibitem 
 {cthem2} R.~C.~Myers and A.~Sinha,
  ``Holographic c-theorems in arbitrary dimensions,''
  arXiv:1011.5819 [hep-th].

\bibitem 
 {casini1} H.~Casini and M.~Huerta,
  ``Entanglement entropy for the n-sphere,''
  Phys.\ Lett.\  B {\bf 694}, 167 (2010)
  [arXiv:1007.1813 [hep-th]].

\bibitem 
 {candow} P.~Candelas and J.~S.~Dowker,
  ``Field Theories On Conformally Related Space-Times: Some Global
  Considerations,''
  Phys.\ Rev.\  D {\bf 19}, 2902 (1979).

\bibitem 
 {solo1} S.~N.~Solodukhin,
  ``Entanglement entropy, conformal invariance and extrinsic geometry,''
  Phys.\ Lett.\  B {\bf 665}, 305 (2008)
  [arXiv:0802.3117 [hep-th]].

\bibitem 
 {duff} See, for example:\\
  M.~J.~Duff,
  ``Observations on Conformal Anomalies,''
  Nucl.\ Phys.\  {\bf B125}, 334 (1977);\\
  M.~J.~Duff,
  ``Twenty years of the Weyl anomaly,''
  Class.\ Quant.\ Grav.\  {\bf 11}, 1387-1404 (1994)
  [hep-th/9308075];\\
  S.~Deser and A.~Schwimmer,
  ``Geometric classification of conformal anomalies in arbitrary dimensions,''
  Phys.\ Lett.\  {\bf B309}, 279-284 (1993)
  [hep-th/9302047].

\bibitem 
 {stud3} J.~S.~Dowker,
  ``Entanglement entropy for even spheres,''
  arXiv:1009.3854 [hep-th].

\bibitem 
 {dowker1} J.~S.~Dowker,
  ``Hyperspherical entanglement entropy,''
  J.\ Phys.\ A {\bf A43}, 445402 (2010)
  [arXiv:1007.3865 [hep-th]].

\bibitem 
 {solo2} S.~N.~Solodukhin,
  ``Entanglement entropy of round spheres,''
  Phys.\ Lett.\  {\bf B693}, 605-608 (2010).
  [arXiv:1008.4314 [hep-th]].

\bibitem 
  {dowker2} J.~S.~Dowker,
  ``Entanglement entropy for odd spheres,''
    [arXiv:1012.1548 [hep-th]].

\bibitem 
 {haag} R.~Haag,
 ``{\sl Local quantum physics: Fields, particles, algebras}'',
  Berlin, Germany: Springer (1992) (Texts and monographs in physics).

\bibitem 
 {cmt} See, for example:\\
H.~Li and F.~D.~M.~Haldane, ``Entanglement Spectrum as a Generalization
of Entanglement Entropy: Identification of Topological Order in
Non-Abelian Fractional Quantum Hall Effect States,'' Phys.\ Rev.\
Lett.\ {\bf 101}, 010504 (2008) [arXiv:0805.0332
[cond-mat.mes-hall]];\\
P.~Calabrese and A.~Lefevre, ``Entanglement spectrum in one-dimensional
systems,'' Phys.\ Rev.\ A {\bf 78}, 032329 (2008);\\
A.~M.~Turner, F.~Pollmann and E.~Berg, ``Topological Phases of
One-Dimensional Fermions: An Entanglement Point of View,'' Phys.\ Rev.\
B {\bf 83}, 075102 (2011) [arXiv:1008.4346 [cond-mat.str-el]];\\
L.~Fidkowski, ``Entanglement spectrum of topological insulators and
superconductors,'' Phys.\ Rev.\ Lett.\ {\bf 104}, 130502 (2010)
[arXiv:0909.2654 [cond-mat.str-el]];\\
H.~Yao and X.-L.~Qi, ``Entanglement entropy and entanglement spectrum
of the Kitaev model,'' 	Phys.\ Rev.\ Lett.\ {\bf 105}, 080501 (2010)
[arXiv:1001.1165 [cond-mat.str-el]].

\bibitem 
 {bisognano} J.~J.~Bisognano and E.~H.~Wichmann,
  ``On The Duality Condition For Quantum Fields,''
  J.\ Math.\ Phys.\  {\bf 17}, 303 (1976);\\
  J.~J.~Bisognano and E.~H.~Wichmann,
  ``On The Duality Condition For A Hermitian Scalar Field,''
  J.\ Math.\ Phys.\  {\bf 16}, 985 (1975).

\bibitem 
 {Unruh} W.~G.~Unruh,
  ``Notes on black hole evaporation,''
  Phys.\ Rev.\  D {\bf 14}, 870 (1976).

\bibitem 
 {Hislop} P.~D.~Hislop and R.~Longo,
  ``Modular Structure Of The Local Algebras Associated With The Free Massless
  Scalar Field Theory,''
  Commun.\ Math.\ Phys.\  {\bf 84}, 71 (1982);

\bibitem 
 {ch1} H.~Casini, M.~Huerta,
  ``Remarks on the entanglement entropy for disconnected regions,''
  JHEP {\bf 0903}, 048 (2009).
  [arXiv:0812.1773 [hep-th]].

\bibitem 
 {roberto} R.~Emparan, ``AdS/CFT duals of topological black
holes and the entropy of zero-energy states,''
  JHEP {\bf 9906}, 036 (1999)
  [arXiv:hep-th/9906040].

\bibitem 
 {topbh} See, for example:\\
S.~Aminneborg, I.~Bengtsson, S.~Holst and P.~Peldan,
  ``Making Anti-de Sitter Black Holes,''
  Class.\ Quant.\ Grav.\  {\bf 13}, 2707 (1996)
  [arXiv:gr-qc/9604005];\\
D.~R.~Brill, J.~Louko and P.~Peldan,
  ``Thermodynamics of (3+1)-dimensional black holes with toroidal or higher
  genus horizons,''
  Phys.\ Rev.\  D {\bf 56}, 3600 (1997)
  [arXiv:gr-qc/9705012];\\
L.~Vanzo,
  ``Black holes with unusual topology,''
  Phys.\ Rev.\  D {\bf 56}, 6475 (1997)
  [arXiv:gr-qc/9705004];\\
 R.~B.~Mann,
  ``Pair production of topological anti-de Sitter black holes,''
  Class.\ Quant.\ Grav.\  {\bf 14}, L109 (1997)
  [arXiv:gr-qc/9607071];\\
D.~Birmingham,
  ``Topological black holes in anti-de Sitter space,''
  Class.\ Quant.\ Grav.\  {\bf 16}, 1197 (1999)
  [arXiv:hep-th/9808032].
R.~Emparan,
  ``AdS membranes wrapped on surfaces of arbitrary genus,''
  Phys.\ Lett.\  B {\bf 432}, 74 (1998)
  [arXiv:hep-th/9804031].

\bibitem 
 {stringx} See, for example:\\
D.~J.~Gross and J.~H.~Sloan,
  ``The Quartic Effective Action for the Heterotic String,''
  Nucl.\ Phys.\  B {\bf 291}, 41 (1987);\\
C.~G.~Callan, E.~J.~Martinec, M.~J.~Perry and D.~Friedan,
  ``Strings In Background Fields,''
  Nucl.\ Phys.\  B {\bf 262}, 593 (1985).

\bibitem 
 {holo} See, for example:\\
R.~C.~Myers, M.~F.~Paulos and A.~Sinha,
  ``Holographic studies of quasi-topological gravity,''
  JHEP {\bf 1008}, 035 (2010)
  [arXiv:1004.2055 [hep-th]];\\
A.~Buchel, J.~Escobedo, R.~C.~Myers, M.~F.~Paulos, A.~Sinha and
M.~Smolkin,
  ``Holographic GB gravity in arbitrary dimensions,''
  JHEP {\bf 1003}, 111 (2010)
  [arXiv:0911.4257 [hep-th]];\\
J.~de Boer, M.~Kulaxizi and A.~Parnachev,
  ``AdS$_7$/CFT$_6$, Gauss-Bonnet Gravity, and Viscosity Bound,''
  JHEP {\bf 1003}, 087 (2010)
  [arXiv:0910.5347 [hep-th]];\\
X.~O.~Camanho and J.~D.~Edelstein,
  ``Causality constraints in AdS/CFT from conformal collider physics and
  Gauss-Bonnet gravity,''
  JHEP {\bf 1004}, 007 (2010)
  [arXiv:0911.3160 [hep-th]];\\
J.~de Boer, M.~Kulaxizi and A.~Parnachev,
  ``Holographic Lovelock Gravities and Black Holes,''
  JHEP {\bf 1006}, 008 (2010)
  [arXiv:0912.1877 [hep-th]];\\
X.~O.~Camanho and J.~D.~Edelstein,
  ``Causality in AdS/CFT and Lovelock theory,''
  JHEP {\bf 1006}, 099 (2010)
  [arXiv:0912.1944 [hep-th]];\\
X.~O.~Camanho, J.~D.~Edelstein and M.~F.~Paulos,
  ``Lovelock theories, holography and the fate of the viscosity bound,''
  arXiv:1010.1682 [hep-th].

\bibitem 
 {lovel} D.~Lovelock,
  ``The Einstein tensor and its generalizations,''
  J.\ Math.\ Phys.\  {\bf 12}, 498 (1971);\\
D.~Lovelock, ``Divergence-free tensorial concomitants,'' Aequationes
Math. {\bf 4}, 127 (1970).

\bibitem 
 {old1} R.~C.~Myers and B.~Robinson,
  ``Black Holes in Quasi-topological Gravity,''
  JHEP {\bf 1008}, 067 (2010)
  [arXiv:1003.5357 [gr-qc]].

\bibitem 
 {WaldEnt} R.~M.~Wald, ``Black hole entropy is the
    Noether charge,''
  Phys.\ Rev.\  D {\bf 48}, 3427 (1993)
  [arXiv:gr-qc/9307038];\\
T.~Jacobson, G.~Kang and R.~C.~Myers,
  ``On Black Hole Entropy,''
  Phys.\ Rev.\  D {\bf 49}, 6587 (1994)
  [arXiv:gr-qc/9312023];\\
V.~Iyer and R.~M.~Wald, ``Some properties of Noether charge and a
proposal for dynamical black hole entropy,''
  Phys.\ Rev.\  D {\bf 50}, 846 (1994)
  [arXiv:gr-qc/9403028].

\bibitem 
 {birrell} N.~D.~Birrell, P.~C.~W.~Davies,
  ``Quantum Fields In Curved Space,''
  Cambridge, Uk: Univ. Pr. ( 1982) 340p.

\bibitem 
 {laflame} R.~Laflamme,
  ``Geometry And Thermofields,''
  Nucl.\ Phys.\  {\bf B324}, 233 (1989).

\bibitem 
 {fursaev} D.~V.~Fursaev, G.~Miele,
  ``Finite temperature scalar field theory in static de Sitter space,''
  Phys.\ Rev.\  {\bf D49}, 987-998 (1994).
  [hep-th/9302078].

\bibitem 
 {vassi} See, for example:\\
 D.~V.~Vassilevich,
  ``Heat kernel expansion: User's manual,''
  Phys.\ Rept.\  {\bf 388}, 279 (2003)
  [arXiv:hep-th/0306138].

\bibitem 
 {bdc} L.~S.~Brown and J.~P.~Cassidy,
  Phys.\ Rev.\  D {\bf 16}, 1712 (1977);\\
T.~S.~Bunch and P.~C.~W.~Davies, ``Quantum Field Theory In De Sitter
Space: Renormalization By Point Splitting,''
  Proc.\ Roy.\ Soc.\ Lond.\  A {\bf 360}, 117 (1978).

\bibitem 
 {friends2} J.~de Boer, M.~Kulaxizi and A.~Parnachev,
  ``Holographic Entanglement Entropy in Lovelock Gravities,''
  arXiv:1101.5781 [hep-th].

\bibitem 
 {jimt} J.~T.~Liu, W.~Sabra and Z.~Zhao,
  ``Holographic c-theorems and higher derivative gravity,''
  arXiv:1012.3382 [hep-th].

\bibitem 
 {aninda6} A.~Sinha,
  ``On higher derivative gravity, c-theorems and cosmology,''
  arXiv:1008.4315 [hep-th].

\bibitem 
 {miguel99x} M.~F.~Paulos,
``Holographic phase space: $c$-functions and black holes as
renormalization group flows,''
  arXiv:1101.5993 [hep-th].

\bibitem 
 {lenny} L.~Susskind and E.~Witten,
  ``The holographic bound in anti-de Sitter space,''
  arXiv:hep-th/9805114.

\bibitem 
 {finn} C.~Holzhey, F.~Larsen and F.~Wilczek,
  ``Geometric and renormalized entropy in conformal field theory,''
  Nucl.\ Phys.\  B {\bf 424}, 443 (1994)
  [arXiv:hep-th/9403108].

\bibitem 
 {adam} A.~Schwimmer and S.~Theisen, ``Entanglement Entropy,
Trace Anomalies and Holography,''
  Nucl.\ Phys.\  B {\bf 801}, 1 (2008)
  [arXiv:0802.1017 [hep-th]].

\bibitem 
 {casini55} H.~Casini,
  ``Mutual information challenges entropy bounds,''
  Class.\ Quant.\ Grav.\  {\bf 24}, 1293 (2007)
  [arXiv:gr-qc/0609126];\\
  H.~Casini and M.~Huerta,
  ``Remarks on the entanglement entropy for disconnected regions,''
  JHEP {\bf 0903}, 048 (2009)
  [arXiv:0812.1773 [hep-th]];\\
H.~Casini and M.~Huerta,
  ``Entanglement entropy in free quantum field theory,''
  J.\ Phys.\ A  {\bf 42}, 504007 (2009)
  [arXiv:0905.2562 [hep-th]].

\bibitem 
 {swingle} B.~Swingle, ``Mutual information and the structure
of entanglement in quantum field theory,''
  arXiv:1010.4038 [quant-ph].

\bibitem 
 {jaff} D.~L.~Jafferis,
  ``The Exact Superconformal R-Symmetry Extremizes Z,''
  arXiv:1012.3210 [hep-th].

\bibitem 
 {ken} K.~A.~Intriligator and B.~Wecht,
  ``The exact superconformal R-symmetry maximizes a,''
  Nucl.\ Phys.\  B {\bf 667}, 183 (2003)
  [arXiv:hep-th/0304128].

\bibitem 
 {ken2} E.~Barnes, K.~A.~Intriligator, B.~Wecht and J.~Wright,
  ``Evidence for the strongest version of the 4d a-theorem, via  a-maximization
  along RG flows,''
  Nucl.\ Phys.\  B {\bf 702}, 131 (2004)
  [arXiv:hep-th/0408156].

\bibitem 
 {curv4} D.~J.~Gross and E.~Witten,
 ``Superstring Modifications Of Einstein's Equations,''
  Nucl.\ Phys.\  B {\bf 277}, 1 (1986);\\
M.~T.~Grisaru, A.~E.~M.~van de Ven and D.~Zanon, ``Four Loop Beta
Function For The N=1 And N=2 Supersymmetric Nonlinear Sigma Model In
Two-Dimensions,''
  Phys.\ Lett.\  B {\bf 173}, 423 (1986).

\bibitem 
 {miguel} M.~F.~Paulos,
  ``Higher derivative terms including the Ramond-Ramond five-form,''
  JHEP {\bf 0810}, 047 (2008)
  [arXiv:0804.0763 [hep-th]].

\bibitem 
 {univres} A.~Buchel, R.~C.~Myers, M.~F.~Paulos and A.~Sinha,
  ``Universal holographic hydrodynamics at finite coupling,''
  Phys.\ Lett.\  B {\bf 669}, 364 (2008)
  [arXiv:0808.1837 [hep-th]].

\bibitem 
 {order} See, for example:\\
M.~B.~Green and M.~Gutperle,
  ``Effects of D-instantons,''
  Nucl.\ Phys.\  B {\bf 498}, 195 (1997)
  [arXiv:hep-th/9701093];\\
 R.~C.~Myers, M.~F.~Paulos and A.~Sinha,
  ``Quantum corrections to eta/s,''
  Phys.\ Rev.\  D {\bf 79}, 041901 (2009)
  [arXiv:0806.2156 [hep-th]].

\bibitem 
 {nochange} D.~Anselmi, D.~Z.~Freedman, M.~T.~Grisaru and A.~A.~Johansen,
  ``Universality of the operator product expansions of SCFT(4),''
  Phys.\ Lett.\  B {\bf 394}, 329 (1997)
  [arXiv:hep-th/9608125];\\
D.~Anselmi, D.~Z.~Freedman, M.~T.~Grisaru and A.~A.~Johansen,
``Nonperturbative formulas for central functions of supersymmetric
gauge theories,''
  Nucl.\ Phys.\  B {\bf 526}, 543 (1998)
  [arXiv:hep-th/9708042].

\bibitem 
 {numerical} R.~Lohmayer, H.~Neuberger, A.~Schwimmer and S.~Theisen,
  ``Numerical determination of entanglement entropy for a sphere,''
  Phys.\ Lett.\  B {\bf 685}, 222 (2010)
  [arXiv:0911.4283 [hep-lat]].

\bibitem 
 {alex} A.~Buchel, R.~C.~Myers and A.~Sinha,
  ``Beyond $\eta/s = 1/4\pi$,''
  JHEP {\bf 0903}, 084 (2009)
  [arXiv:0812.2521 [hep-th]];\\
Y.~Kats and P.~Petrov, ``Effect of curvature squared corrections in AdS
on the viscosity of the dual gauge theory,''
  JHEP {\bf 0901}, 044 (2009)
  [arXiv:0712.0743 [hep-th]].

\bibitem 
 {ddbrane} C.~P.~Bachas, P.~Bain and M.~B.~Green,
  ``Curvature terms in D-brane actions and their M-theory origin,''
  JHEP {\bf 9905}, 011 (1999)
  [arXiv:hep-th/9903210].

\bibitem 
 {juanbh} J.~M.~Maldacena,
  ``Eternal black holes in Anti-de-Sitter,''
  JHEP {\bf 0304}, 021 (2003)
  [arXiv:hep-th/0106112].

\bibitem 
 {rotate} See, for example:\\
 D.~Klemm, V.~Moretti and L.~Vanzo,
  ``Rotating topological black holes,''
  Phys.\ Rev.\  D {\bf 57}, 6127 (1998)
  [Erratum-ibid.\  D {\bf 60}, 109902 (1999)]
  [arXiv:gr-qc/9710123];\\
M.~H.~Dehghani, ``Rotating topological black holes in various
dimensions and AdS/CFT correspondence,''
  Phys.\ Rev.\  D {\bf 65}, 124002 (2002)
  [arXiv:hep-th/0203049];\\
R.~B.~Mann, ``Charged topological black hole pair creation,''
  Nucl.\ Phys.\  B {\bf 516}, 357 (1998)
  [arXiv:hep-th/9705223];\\
R.~G.~Cai and A.~Wang, ``Thermodynamics and stability of hyperbolic
charged black holes,''
  Phys.\ Rev.\  D {\bf 70}, 064013 (2004)
  [arXiv:hep-th/0406057].


\end{thebibliography}
\end{document}